# Emergence of interfacial magnetism in strongly-correlated nickelate-titanate superlattices

Teguh Citra Asmara[1,2,*], Robert J. Green[3,4], Andreas Suter[5], Yuan Wei[1], Wenliang Zhang[1], Daniel Knez[6], Grant Harris[3], Yi Tseng[1], Tianlun Yu[1], Davide Betto[7], Mirian Garcia-Fernandez[8], Stefano Agrestini[8], Yannick Maximilian Klein[9], Neeraj Kumar[10], Carlos William Galdino[1], Zaher Salman[5], Thomas Prokscha[5], Marisa Medarde[9], Elisabeth Müller[11], Yona Soh[10], Nicholas B. Brookes[7], Ke-Jin Zhou[8], Milan Radovic[1,**], Thorsten Schmitt[1,***]

[1]*PSI Center for Photon Science, Paul Scherrer Institute, CH-5232 Villigen PSI, Switzerland*
[2]*European X-Ray Free-Electron Laser Facility GmbH, 22869 Schenefeld, Germany*
[3]*Department of Physics & Engineering Physics, University of Saskatchewan, Saskatoon, Canada*
[4]*Stewart Blusson Quantum Matter Institute, University of British Columbia, Vancouver, British Columbia, Canada*
[5]*Laboratory for Muon-Spin Spectroscopy, Paul Scherrer Institute, CH-5232 Villigen PSI, Switzerland*
[6]*Institute of Electron Microscopy and Nanoanalysis, Graz University of Technology, Steyergasse 17, 8010 Graz, Austria*
[7]*European Synchrotron Radiation Facility, F-38043 Grenoble Cedex, France*
[8]*Diamond Light Source, Harwell Campus, Didcot OX11 0DE, United Kingdom*
[9]*Laboratory for Multiscale Materials Experiments, Paul Scherrer Institute, CH-5232 Villigen PSI, Switzerland*
[10]*Paul Scherrer Institute, CH-5232 Villigen PSI, Switzerland*
[11]*Electron Microscopy Facility, Paul Scherrer Institute, CH-5232 Villigen PSI, Switzerland*
*teguh.citra.asmara@xfel.eu*
**milan.radovic@psi.ch*
***thorsten.schmitt@psi.ch*

**Strongly-correlated transition-metal oxides are widely known for their various exotic phenomena. This is exemplified by rare-earth nickelates such as LaNiO$_3$, which possess intimate interconnections between their electronic, spin, and lattice degrees of freedom. Their properties can be further enhanced by pairing them in hybrid heterostructures, which can lead to hidden phases and emergent phenomena. An important example is the LaNiO$_3$/LaTiO$_3$ superlattice, where an interlayer electron transfer has been observed from LaTiO$_3$ into LaNiO$_3$ leading to a high-spin state. However, macroscopic emergence of magnetic order associated with this high-spin state has so far not been observed. Here, by using muon spin rotation, x-ray absorption, and resonant inelastic x-ray scattering, we present direct evidence of an emergent antiferromagnetic order with high magnon energy and exchange interactions at the LaNiO$_3$/LaTiO$_3$ interface. As the magnetism is purely interfacial, a single LaNiO$_3$/LaTiO$_3$ interface can essentially behave as an atomically thin strongly-correlated quasi-two-dimensional antiferromagnet, potentially allowing its technological utilisation in advanced spintronic devices. Furthermore, its strong quasi-two-dimensional magnetic correlations, orbitally-polarized planar ligand holes, and layered superlattice design make its electronic, magnetic, and lattice configurations resemble the precursor states of superconducting cuprates and nickelates, but with an $S \rightarrow 1$ spin state instead.**

## 1. Introduction

The heterostructure interfaces between strongly-correlated transition-metal oxides (TMOs) have become a fertile ground for the discovery of new properties of matter. In these systems, interlayer interactions such as charge transfer, quantum confinement, epitaxial strain, and proximity effect can tune the intertwined charge, orbital, and spin degrees of freedom of their correlated electrons, leading to emergent interfacial phenomena not found in their bulk forms.[1,2] An important example is the interfacial magnetism,[2-8] which has found many technological applications in data storage and spintronic devices.[9] Remarkably, in TMO heterostructures it can emerge even if one or both of the parent materials are non-magnetic, as spin order is often interconnected with lattice and charge orders.[3-8] The quasi-two-dimensionality of interface magnetism is also relevant in unconventional high-temperature superconductivity, as strong quasi-two-dimensional antiferromagnetic correlations are believed to play a key role in the pairing mechanisms of superconducting electrons.[10,11]

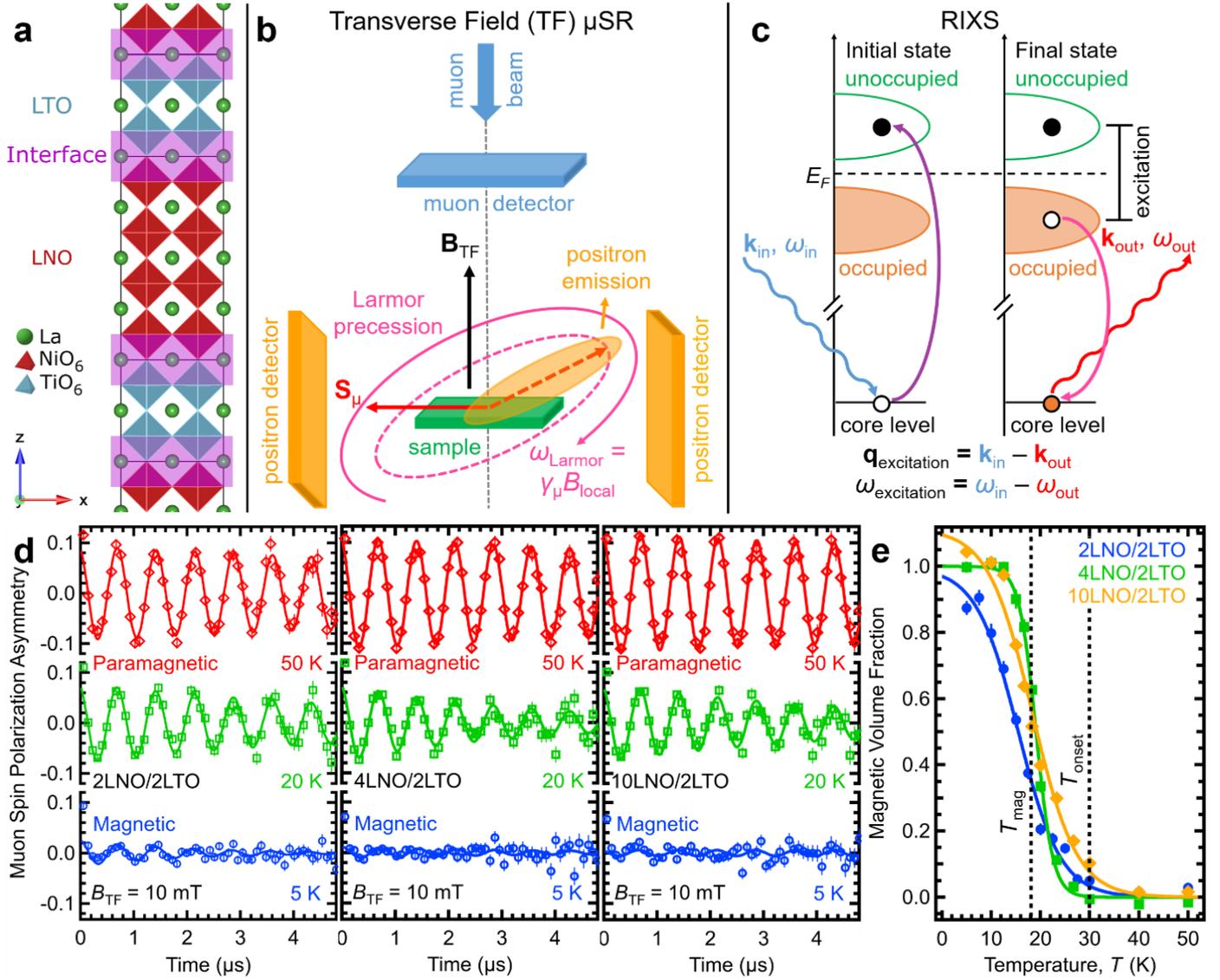

**Figure 1.** Emergent magnetism in LNO/LTO superlattices. (a) Schematics of LNO/LTO superlattice crystal structure. The LNO layer thickness is variable, and here 4LNO/2LTO is shown as a representative example. (b) Schematics of the transverse-field muon spin rotation/relaxation (TF-µSR) experiment. Here, a transverse field, $B_{TF}$, is applied to the sample, causing the muon spins, $S_\mu$, to undergo Larmor precessions with frequencies proportional to the muon gyromagnetic ratio, $\gamma_\mu$, and the local field they experience inside the sample, $B_{local}$. (c) Schematics of the resonant inelastic x-ray scattering (RIXS) experiment. Here, $k_{in}$ ($k_{out}$) and $\omega_{in}$ ($\omega_{out}$) are the incoming (outgoing) momentum and energy of the photons, respectively, while $q_{excitation}$ ($\omega_{excitation}$) is the momentum (energy) transferred into the excitation. (d) and (e), Temperature-dependent muon spin polarization asymmetry in TF-µSR geometry and magnetic volume fraction of the LNO/LTO samples, respectively, showing a sharp magnetic transition below $T_{onset} \approx 30$ K and a transition temperature of $T_{mag} \approx 18$ K. The solid lines are guides to the eye.

Recently, a family of TMOs called the nickelates have reattracted attention due to the discovery of superconductivity in infinite-layer (Nd,Sr)NiO$_2$ (superconducting critical temperature, $T_c \approx 15$ K)[11] and highly-pressurized La$_3$Ni$_2$O$_7$ ($T_c$ = 80 K).[13] In particular, rare-earth nickelates ($R$NiO$_3$), the base compound of (Nd,Sr)NiO$_2$, possess an unusual negative charge-transfer ground state[14-19] where the energy difference between their charge-transfer Ni-3$d^8\underline{L}^1$ state with their nominal Ni-3$d^7$ configuration is negative, causing Ni-3$d^8\underline{L}^1$ to become the preferred ground state with an oxygen ligand hole ($\underline{L}$) dominating the Fermi level. At low temperatures, most $R$NiO$_3$ (except LaNiO$_3$) undergo intertwined phase transitions where this paramagnetic metallic Ni-3$d^8\underline{L}^1$ configuration transforms into a structurally-reconstructed antiferromagnetic insulating phase, where half of the Ni sites become high-spin Ni-3$d^8$ ($S \rightarrow 1$) and the rest transform into low-spin Ni-3$d^8\underline{L}^2$ ($S \rightarrow 0$), indicating an intimate connection between their electronic, spin, and lattice interactions.[14,19-22] In particular, although LaNiO$_3$ (LNO) does not exhibit these interconnected phase transitions like other $R$NiO$_3$, its strongly-correlated metallic carriers are on the verge of localization, making it especially susceptible to external stimuli.[23,24] This can be employed by pairing LNO with other TMOs in hybrid heterostructures, resulting in many remarkable emergent phenomena including confinement-induced magnetism in LNO/LaAlO$_3$,[5] exchange bias in LNO/LaMnO$_3$,[6] and interlayer charge transfer in LNO/LaTiO$_3$ (LNO/LTO).[25-27]

The LNO/LTO superlattice[25-27] (Figure 1a) is especially noteworthy, as its charge transfer can lead to further emergent phenomena. Bare LTO is an antiferromagnetic Mott-insulator with a valence configuration of Ti$^{3+}$-3$d^1$.[28] As the electron affinity of Ti is much lower than Ni, this available Ti-3$d^1$ electron from LTO can easily transfer into LNO[7,25-27] which can readily receive the electron due to its ligand holes. This leads to a high-spin state of Ni-3$d^8$ ($S \rightarrow 1$) in all Ni sites,[29] as indicated by its x-ray absorption spectrum (XAS) which resembles that of NiO,[26,30] the prototypical example of a Ni-3$d^8$ ($S \rightarrow 1$) compound. In comparison, in bare $R$NiO$_3$ (for $R \neq$ La) this high-spin configuration only occurs in half of the Ni sites.[15-20,22] Therefore, the electron transfer can effectively double the Ni magnetic moment.

Furthermore, due to the confinement by LTO layers, the magnetic order resulting from this high-spin state is predicted to also be quasi-two-dimensional[29] (in contrast to the three-dimensional magnetic order of bare $R$NiO$_3$[15-20,22] and NiO[30]), making it potentially suitable for applications in advanced spintronic devices as its reduced dimensionality can make its magnetic properties more controllable by external stimuli.[31] However, so far the macroscopic emergence of magnetic order associated with this high-spin state has not been found in LNO/LTO.

The interlayer charge transfer also produces a large orbital polarization in LNO/LTO[25-27], which is significant because, along with strong quasi-two-dimensional antiferromagnetic correlations, it has been theorized to be a prerequisite for achieving superconductivity in nickelates.[32] This is especially relevant because the orbital polarization in LNO/LTO results in a preferential occupation of ligand holes in the Ni-$3d_{x^2-y^2}$ + O-$2p_{xy}$ hybridized state[27] similar to the normal state of La$_3$Ni$_2$O$_7$ which becomes superconducting under high pressure[13,33] and reminiscent of superconducting cuprates.[10,11] Currently, the two types of nickelate superconducting phases (oxygen-reduced infinite-layer nickelates[12] and highly-pressurized La$_3$Ni$_2$O$_7$[13]) are still difficult to stabilize, either due to their chemical volatility caused by low oxidation state[12] and unforeseen hydrogen contamination,[34] or the need of applying gigapascals of pressure to reach the superconducting state,[13] respectively. Therefore, tuning these LNO-based heterostructures can provide another pathway for achieving nickelate-based superconductivity, especially as the layered structures of these superconducting nickelates can, in principle, be replicated with suitable heterostructure designs.[32] However, LNO/LTO has so far still lacked the second key ingredient of strong quasi-two-dimensional antiferromagnetic correlations, which has hindered the research progress on this front.

Here, by tuning the interlayer charge transfer between LTO and LNO, we present direct evidence of an emergent quasi-two-dimensional $S \to 1$ antiferromagnetic order at the LNO/LTO interface with high magnetic exchange correlations. We sensitively probe the emergence of this order using muon spin rotation[35] (µSR, Figure 1b), which shows that the magnetism is quasi-two-dimensionally confined to the interface. We investigate the microscopic details of the interlayer charge transfer using XAS, x-ray emission spectroscopy (XES), and resonant inelastic x-ray scattering[36] (RIXS, Figure 1c) spectroscopies. Particularly, the high-resolution RIXS results reveal that the magnetic order gives rise to a high and dispersive magnon energy of ~100 meV, implying strong quasi-two-dimensional exchange interactions between the spins.

## 2. Results

The ($n$LNO/$m$LTO)$_{30}$ superlattices are fabricated using pulsed-laser deposition on lattice-matched (110)$_{orthorhombic}$ NdGaO$_3$ substrates to minimize epitaxial strain on LNO. To ensure the stoichiometric growth of both LNO and LTO, N$_2$ is used as the process gas instead of O$_2$, so that the growth requirements of high ambient pressure and low oxygen partial pressure for stoichiometric LNO[27] and LTO,[37] respectively, can both be fulfilled. To examine the effects of varying interlayer electron transfer and quantum confinement on their emergent properties, three samples are prepared: 2LNO/2LTO, 4LNO/2LTO, and 10LNO/2LTO. *A priori*, 2LNO/2LTO is expected to have the highest confinement and amount of electron transfer due to having the thinnest LNO layers and 1:1 LTO-to-LNO ratio, respectively, while 10LNO/2LTO should have the least of both. The 2LNO/2LTO is highly insulating (Figure S2) and the electrical resistivity decreases as LNO thickness increases, with 10LNO/2LTO having a subtle insulator-to-metal transition at ~135 K.

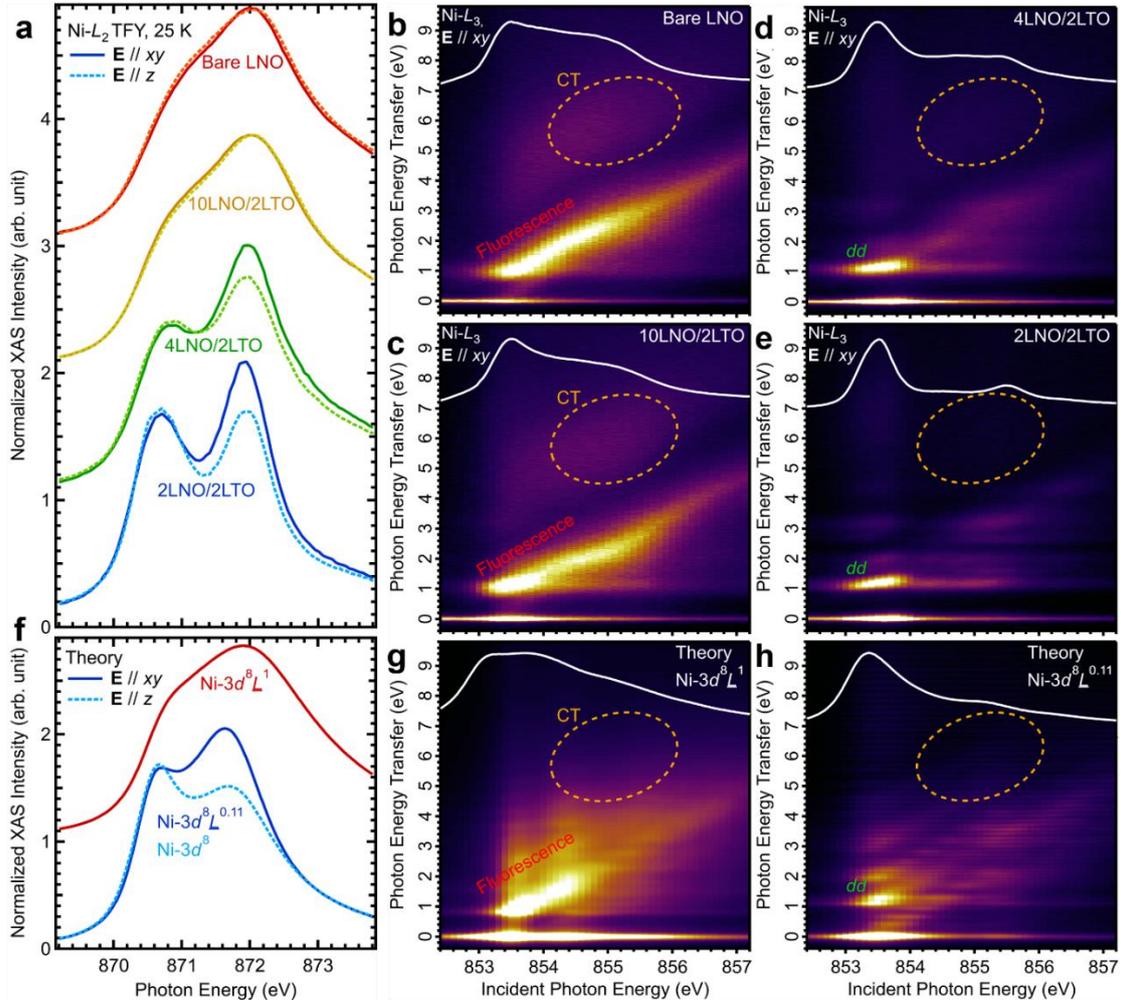

**Figure 2.** Ni-$L_{2,3}$ x-ray spectra of LNO/LTO superlattices. (a) Experimental Ni-$L_2$ x-ray absorption (XAS) spectra of the samples, showing data for both in-plane (**E** // $xy$, straight lines) and out-of-plane (**E** // $z$, dotted lines) polarizations. The 2LNO/2LTO (and, to a lesser extent, 4LNO/2LTO) spectra show a clear x-ray linear dichroism (XLD). The spectra are taken with the total fluorescence yield (TFY) mode and have been self-absorption corrected. (b) – (e) Experimental Ni-$L_3$ resonant inelastic x-ray scattering (RIXS) energy maps of the samples, taken with in-plane (**E** // $xy$) polarization, showing, in particular, the charge-transfer (CT) and fluorescence excitations of bare LNO, which vanish in 2LNO/2LTO. The corresponding Ni-$L_3$ XAS spectra, after the removal of the La-$M_4$ peak, (white lines) are also shown for reference. (f) Theoretical simulations of Ni-$L_2$ XAS spectra obtained using cluster-based calculations, showing the cases for Ni-$3d^8$-like and Ni-$3d^8\underline{L}^1$-like ground states. The experimental XLD can be simulated by introducing an exchange interaction in the $xy$-plane and varying the ligand hole filling, as demonstrated by the difference between the simulated spectra for Ni-$3d^8\underline{L}^{0.11}$ and Ni-$3d^8$ states along the $xy$- and $z$-direction, respectively. (g) and (h), Theoretical simulations of Ni-$L_3$ RIXS energy maps for Ni-$3d^8\underline{L}^1$-like and Ni-$3d^8$-like ground states, respectively, showing the CT and fluorescence excitations that vanish as the ligand hole is filled. The RIXS map in (h) is calculated based on the Ni-$3d^8\underline{L}^{0.11}$ state to better simulate the incomplete electron transfer in the $xy$-plane.

The emergence of magnetism is investigated by μSR.[35] Implanted spin-polarized muons in the sample preferentially decay into positrons along their spin direction, which act as highly sensitive probes for detecting local magnetic moments within the sample (Figure 1b). The μSR experiments are done in a transverse field geometry (TF-μSR) at the Low-Energy Muon Facility of the Swiss Muon Source.[38] In this geometry, undamped oscillations in the time-dependent muon spin polarization asymmetry indicate that the sample is paramagnetic. If the oscillations diminish, it is a direct indication that magnetic order has emerged within the sample.[35] From the temperature-dependent decrease of the oscillation amplitude, the magnetic volume fraction, $f_{mag}$, of the sample can be extracted to determine the magnetic transition critical temperatures. Remarkably, Figure 1d – 1e shows that the TF-μSR oscillations of the three LNO/LTO samples are all quenched below 20 K, and their $f_{mag}$ sharply increases from 0 (paramagnetic state) to 1 (fully-magnetized state). The TF-μSR results thus reveal the emergence of a magnetic phase in LNO/LTO below $T_{onset} \approx 30$ K, with a critical temperature (defined at $f_{mag} = 0.5$) of $T_{mag} \approx 18$ K.

Intriguingly, the magnetic behaviour of 2LNO/2LTO and the other two superlattices with thicker LNO are extremely similar despite their large differences in LNO thickness. This shows that 2 LNO uc thickness is enough to establish the emergent magnetism (as using thicker LNO layers do not change the magnetic transition), implying that the magnetically active region is contained only within these 2 LNO uc. This quasi-two-dimensionality is also supported by μSR data taken at zero-field geometry (Figure S3), which reveals that the internal magnetic field distribution inside the samples is widely uneven: some regions are strongly magnetic, while others are non-magnetic. However, this quasi-two-dimensional magnetism cannot originate only from the geometrical confinement of the LNO layers, as the magnetism persists in 10LNO/2LTO with its thick LNO layers. This is in contrast to LNO/LaAlO$_3$, whose magnetism only exists in strongly-confined 2LNO/2LaAlO$_3$ but disappears in moderately-confined 4LNO/4LaAlO$_3$.[5] Thus, the magnetism in LNO/LTO is much more robust than in LNO/LaAlO$_3$, as it emerges without needing a strict geometrical confinement of the LNO layers.

Therefore, the magnetism most likely comes from the electron transfer from LTO, which can transform LNO into its high-spin Ni-3$d^8$ state.[25-27,29] To investigate this, we probe the samples with XAS, XES, and RIXS using the ADRESS beamline of the Swiss Light Source.[39,40] We first focus on the Ni-$L_{2,3}$ edges to reveal the electronic configuration of the superlattices compared to bare LNO. Figure 2a shows that the Ni-$L_2$ XAS of bare LNO consists of two partially-melded peaks with differing intensities. When LTO is introduced to form the LNO/LTO superlattices, these peaks become more separated and their peak ratio becomes more equal in 2LNO/2LTO. Furthermore, the spectra of 2LNO/2LTO also shows a considerable x-ray linear dichroism (XLD) between the in-plane $xy$- and out-of-plane $z$-polarization, which is not present in bare LNO.

From these results, it is clear that there are significant differences between the ground states of LNO/LTO compared to bare LNO. To investigate the details of how they are different, we use Ni-$L_3$ RIXS to probe specific elementary excitations that are characteristic for different electronic states of LNO[17,36] (Figure 2b – 2e). Particularly, the charge-transfer (CT) and itinerant fluorescence excitations prominent in bare LNO come from the ligand hole and hence are the fingerprints for its Ni-$3d^8\underline{L}^1$ state.[17] When LTO is introduced, these excitations gradually diminish until they almost vanish in 2LNO/2LTO, signifying that the ligand hole is almost completely filled in 2LNO/2LTO due to the pairing with LTO. The $dd$ excitations thus become the most prominent features due to the localized nature of 2LNO/2LTO.

To support the analysis, we perform single impurity Anderson model (SIAM) calculations[18,41-43] to obtain theoretical simulations of the LNO spectra under different electron doping conditions (Figure 2f – 2h). Indeed, the calculations show that the Ni-$L_2$ XAS spectra of bare LNO resemble that of the Ni-$3d^8\underline{L}^1$ state, while 2LNO/2LTO resembles Ni-$3d^8$-like state similar to NiO.[26,30] The theoretical Ni-$L_3$ RIXS spectra of Ni-$3d^8\underline{L}^1$ also shows a prominent CT excitation indicative of a ligand hole and itinerant fluorescence, while in the Ni-$3d^8$-like state both the CT band and the itinerant fluorescence are suppressed consistent with previous results.[27] Thus, the electron transfer from LTO has preferentially filled the ligand holes of LNO, transforming the LNO ground state from its initial Ni-$3d^8\underline{L}^1$ state[11-16] into a Ni-$3d^8$-like state, which is in the positive charge-transfer regime.[44]

Meanwhile, the strong XLD in the Ni-$L_2$ XAS of 2LNO/2LTO (Figure 2a) can originate from three main mechanisms: a strain-induced crystal field distortion, different degrees of in-plane versus out-of-plane electron transfer, and a magnetic linear dichroism effect caused by the emergent magnetic order. While all three mechanisms can be present, we find that this XLD can be reproduced in the SIAM calculation using a combination of asymmetric electron transfer (where the electron transfer along the $z$-direction is more complete than the $xy$-plane) and the inclusion of a 120 meV exchange field in the $xy$-plane (Figure 2f). As neither effect can fully capture the dichroism by itself, this XLD is evidence of both (1) the lifting of degeneracy due to the emergence of magnetic correlations among the high-spin Ni-$3d^8$-like sites and (2) an anisotropy in the electron transfer path, as it is more complete along the $z$-direction (resulting in the hole-less Ni-$3d^8$ state) and less complete in the $xy$-plane (resulting in the Ni-$3d^8\underline{L}^{0.11}$ state). Consequently, this means the LNO layers in 2LNO/2LTO contain residual, singly-polarized ligand holes that selectively inhabit only the planar Ni-$3d_{x^2-y^2} + \text{O-}2p_{xy}$ hybridized band, consistent with the large orbital polarization and XLD spectra observed previously.[25-27]

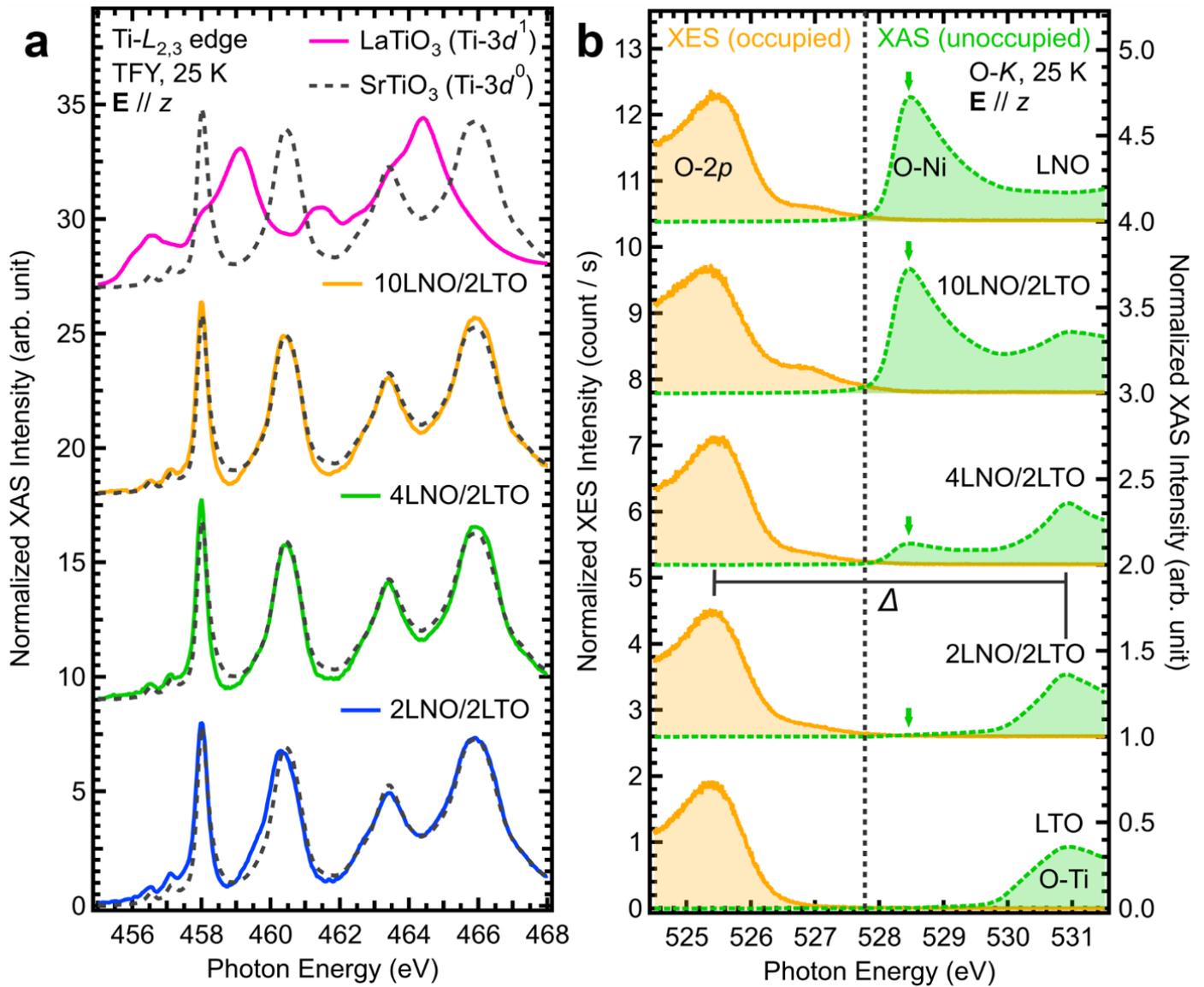

**Figure 3.** Ti-$L_{2,3}$ and O-$K$ x-ray spectra of LNO/LTO superlattices. (a) Experimental Ti-$L_{2,3}$ x-ray absorption spectroscopy (XAS) of the samples for out-of-plane (**E** // $z$) polarization. The Ti-$L_{2,3}$ XAS spectra of LNO/LTO resemble the diamagnetic Ti$^{4+}$-3$d^0$ configuration of SrTiO$_3$ instead of the antiferromagnetic Ti$^{3+}$-3$d^1$ configuration of bare LTO, signifying that the Ti-3$d$ orbitals of LTO in LNO/LTO have indeed been emptied by electron transfer into the LNO layers. The Ti-$L_{2,3}$ XAS spectrum of bare LTO is reprinted with permission from M. Haverkort *et al.*, *Phys. Rev. Lett* **94**, 056401 (2005)[28]. Copyright (2005) by the American Physical Society. DOI: https://doi.org/10.1103/PhysRevLett.94.056401. (b) Experimental O-$K$ x-ray emission spectroscopy (XES) and XAS of the samples (equivalent to occupied and unoccupied density of states (DOS), respectively) for **E** // $z$ polarization, showing the diminishing of the O-Ni hybridized peak until a positive charge-transfer gap of $\Delta \approx 5$ eV opens for 2LNO/2LTO. The XAS data are taken with the total fluorescence yield (TFY) mode and have been self-absorption corrected, while the O-$K$ XES data are obtained by taking non-resonant inelastic x-ray scattering spectra at ~5.5 eV above the O-Ni peak.

To verify the electron transfer from the LTO side, the Ti-$L_3$ XAS spectra of the samples (Figure 3a and Figure S5a) show that the LTO layers of LNO/LTO all resemble that of $SrTiO_3$ with its diamagnetic Ti-$3d^0$ configuration, in contrast to bare LTO whose ground state is antiferromagnetic Ti-$3d^1$.[28] Therefore, the Ti-$3d$ orbitals of the LTO layers have indeed been emptied and their electrons all transferred into the LNO layers, consistent with previous results.[26] Furthermore, we also probe the electron transfer from the ligand side by comparing the O-$K$ XES and XAS spectra of the samples (equivalent to occupied and unoccupied density of states, respectively) in Figure 3b and Figure S5b. The Ni-$3d^8\underline{L}^1$ state of bare LNO is characterized by a distinct O-Ni hybridized XAS peak coming from the excitation into the unoccupied O ligand hole.[45] As the LNO thickness decreases, this O-Ni peak diminishes and eventually disappears, indicating the (almost) complete annihilation of the ligand hole in 2LNO/2LTO (which previously was only achieved partially[26]) and resulting in a positive charge-transfer gap of $\Delta \approx 5$ eV.

These results comprehensively show that the magnetism in LNO/LTO originates from electron transfer from LTO, which quenches the LNO ligand holes and transforms it into the high-spin Ni-$3d^8$ ($S \rightarrow 1$) state with minimal interference from the ligand. At first, it may seem that only 2LNO/2LTO (and to a lesser extent, 4LNO/2LTO) undergoes this transformation, as the spectra for 10LNO/2LTO still resemble those of bare LNO despite having the same magnetic behaviour as 2LNO/2LTO. This can be resolved by considering that, based on µSR results (Figure 1d – 1e), the emergent magnetism of LNO/LTO is contained within a small region ~2 LNO uc thick, indicating that the electron transfer only happens within this small region around the LNO/LTO interface. If the transferred electrons would instead be distributed uniformly throughout the whole LNO layer, each LNO uc in 10LNO/2LTO would only receive 0.2 electron from LTO (as each LTO uc can only donate 1 electron), resulting in a ground state of Ni-$3d^8\underline{L}^{0.8}$. This would lead to a much smaller $\mu_{Ni}$ and ultimately a different magnetic behaviour compared to 2LNO/2LTO, in contrast with the µSR data. Thus, to be consistent with both the muon and x-ray spectroscopy results, the electron transfer must be received only by the interfacial LNO uc (leading to their transformation to the high-spin Ni-$3d^8$ state), while LNO layers far away from the interface receive almost no electrons and stay unchanged in their initial paramagnetic Ni-$3d^8\underline{L}^1$-like state (Figure 4). Within this context, the x-ray spectra of 10LNO/2LTO look similar to bare LNO simply because 10LNO/2LTO has many non-interfacial LNO unit cells, which overwhelm its high-spin interface state. Cross-sectional scanning transmission electron microscopy – electron energy loss spectra (STEM-EELS) of 10LNO/2LTO (Figure S9) also support this analysis, as the O-Ni peak of the LNO layer is suppressed near the interface (indicating a high-spin Ni-$3d^8$-like state) but remains strong away from the interface (indicating a bare-LNO-like Ni-$3d^8\underline{L}^1$ state). This shows that that the charge transfer and thus the magnetism are indeed confined at the interfaces of LNO/LTO.

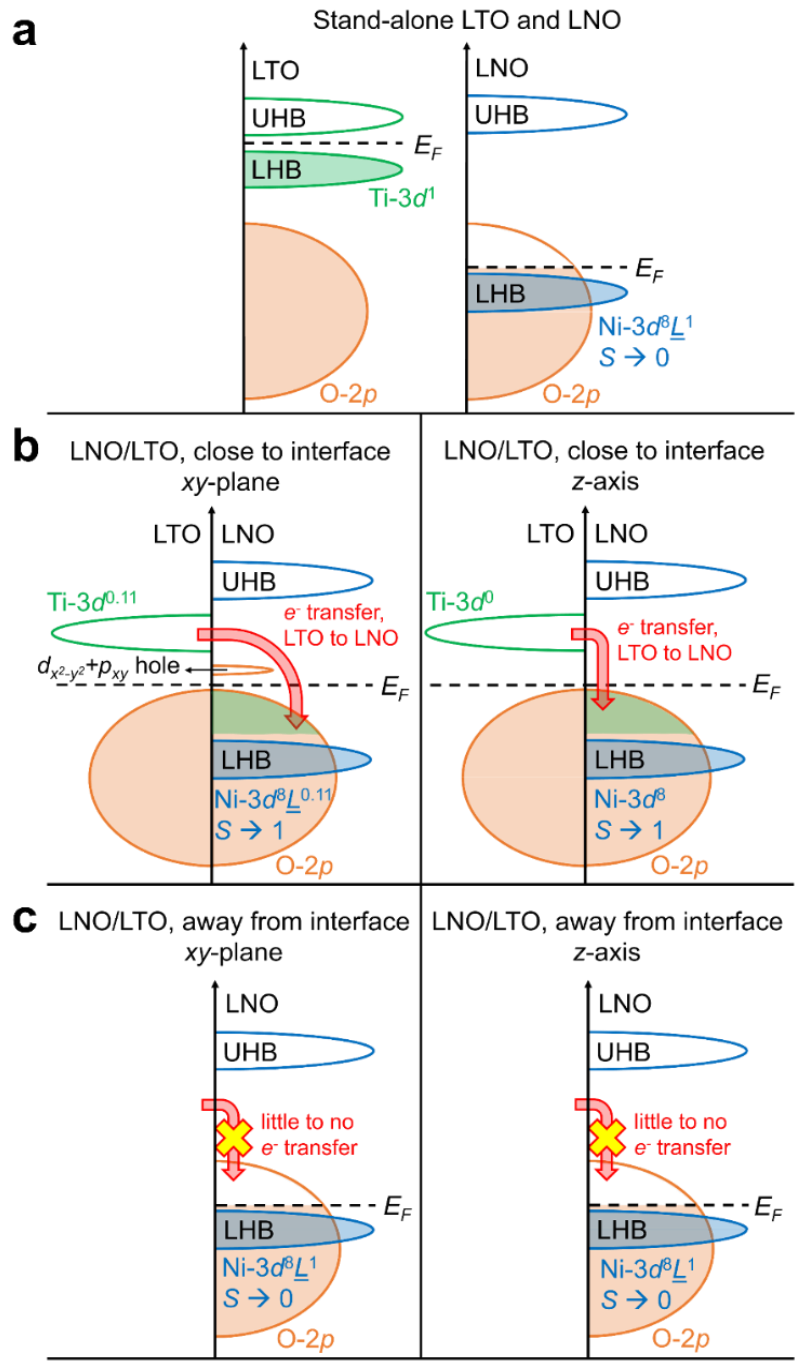

**Figure 4.** Schematics of the interlayer electron ($e^-$) transfer from LTO to LNO in LNO/LTO superlattices. (a) Electronic band structures of bare LNO and bare LTO. The UHB and LHB denote the upper and lower Hubbard band, respectively. (b) Electrons from the LTO Ti-3$d$ orbital move across the Fermi level, $E_F$, to fill the ligand hole of LNO, transforming the LNO ground state from the paramagnetic Ni-3$d^8\underline{L}^1$-like to a magnetic ($S \to 1$) Ni-3$d^8$-like state. The electron transfer processes are concentrated near the LNO/LTO interfaces, so only interfacial LNO unit cells become magnetic. There is also a distinct dichroism, as the electron transfer along the out-of-plane $z$-axis is more complete than along the $xy$-plane, resulting in singly-polarized residual holes in the planar Ni-3$d_{x^2-y^2}$ + O-2$p_{xy}$ hybridized band. (c) There is little to no electron transfer away from the interfaces, so non-interfacial LNO unit cells stay in their initial paramagnetic Ni-3$d^8\underline{L}^1$-like state.

Therefore, by combining muon, x-ray, and cross-sectional electron spectroscopies on LNO/LTO with different LNO layer thicknesses, we have discovered the emergence of an interfacial quasi-two-dimensional magnetism in LNO/LTO superlattices. To further investigate its microscopic details, we use high-resolution RIXS (FWHM *ca.* 40 meV) to probe the magnon excitations associated with this magnetic order. As the magnon energy directly correlates to the magnetic exchange coupling,[22,30,46-49] this allows us to compare the magnetic correlation strength of LNO/LTO with other relevant compounds such as other nickelates and the cuprate superconductors. The experiments are performed at the Ni-$L_2$ edge using the ID32 beamline of the European Synchrotron Radiation Facility[50] and the I21 beamline of the Diamond Light Source.[51] The results (Figure 5a) reveal the presence of significant spectral weight at low energy transfer, which can be well-fitted by a dominant magnon excitation (Figure 5b) that shows a slight but noticeable dispersion along both the $(1,0,0)_{pc=pseudocubic}$ and $(1,1,0)_{pc}$ directions (Figure 5c).

As the spectra are taken at 20 – 30 K (slightly above $T_{mag}$) due to beamline limitations, the magnon is underdamped with a (fitted) damping coefficient of ~80 meV and modulated by the Bose-Einstein distribution. These cause the apparent central energy of the magnon, $\omega_{cen} \approx 75$ meV, to be lower than its actual natural/undamped energy, $\omega_{mag}$. To extract $\omega_{mag}$, we fit the magnon with a damped harmonic oscillator model[51] which results in a momentum-dispersed $\omega_{mag}$ of ~90 – 105 meV that increases with the in-plane momentum transfer (Figure 5d). This $\omega_{mag}$ is ~3 times higher than that of its parent material $R$NiO$_3$ ($\omega_{mag} \approx 20 – 40$ meV)[22] and almost the same as that of La$_2$NiO$_4$ ($\omega_{mag} \approx 100$ meV) which has similar planar Ni-3$d^8$ ($S = 1$) magnetic correlations.[48] Importantly, this $\omega_{mag}$ also approaches the typical values for superconducting cuprates ($\omega_{mag} \approx 200 – 300$ meV)[46,47,49] and NdNiO$_2$ ($\omega_{mag} \approx 150 – 200$ meV),[53] the parent compound of superconducting (Nd,Sr)NiO$_2$.[12] Considering that some theories[10] and experimental observations[47,49,54,55] have suggested a positive correlation between (para)magnon and magnetic exchange energies with superconducting gap and $T_c$ in cuprates, this high $\omega_{mag}$ indicates that LNO/LTO possesses strong magnetic exchange correlations that mimic the precursor states of superconducting nickelates and cuprates, especially with its quasi-two-dimensional magnetic order.

The intensity of these magnon excitations decrease as LNO thickness increases (Figure 5e) because the fraction of magnetic interfacial LNO uc in the superlattice also decreases, but even in 10LNO/2LTO it is still twice the intensity of the low-energy spectral weight of bare LNO. Interestingly, the damped magnon still persists at higher temperatures far above the magnetic $T_{onset}$, indicating that locally the magnetic exchange coupling of LNO/LTO is still strong enough to retain significant short-range magnetic correlations at higher temperatures consistent with its high $\omega_{mag}$.

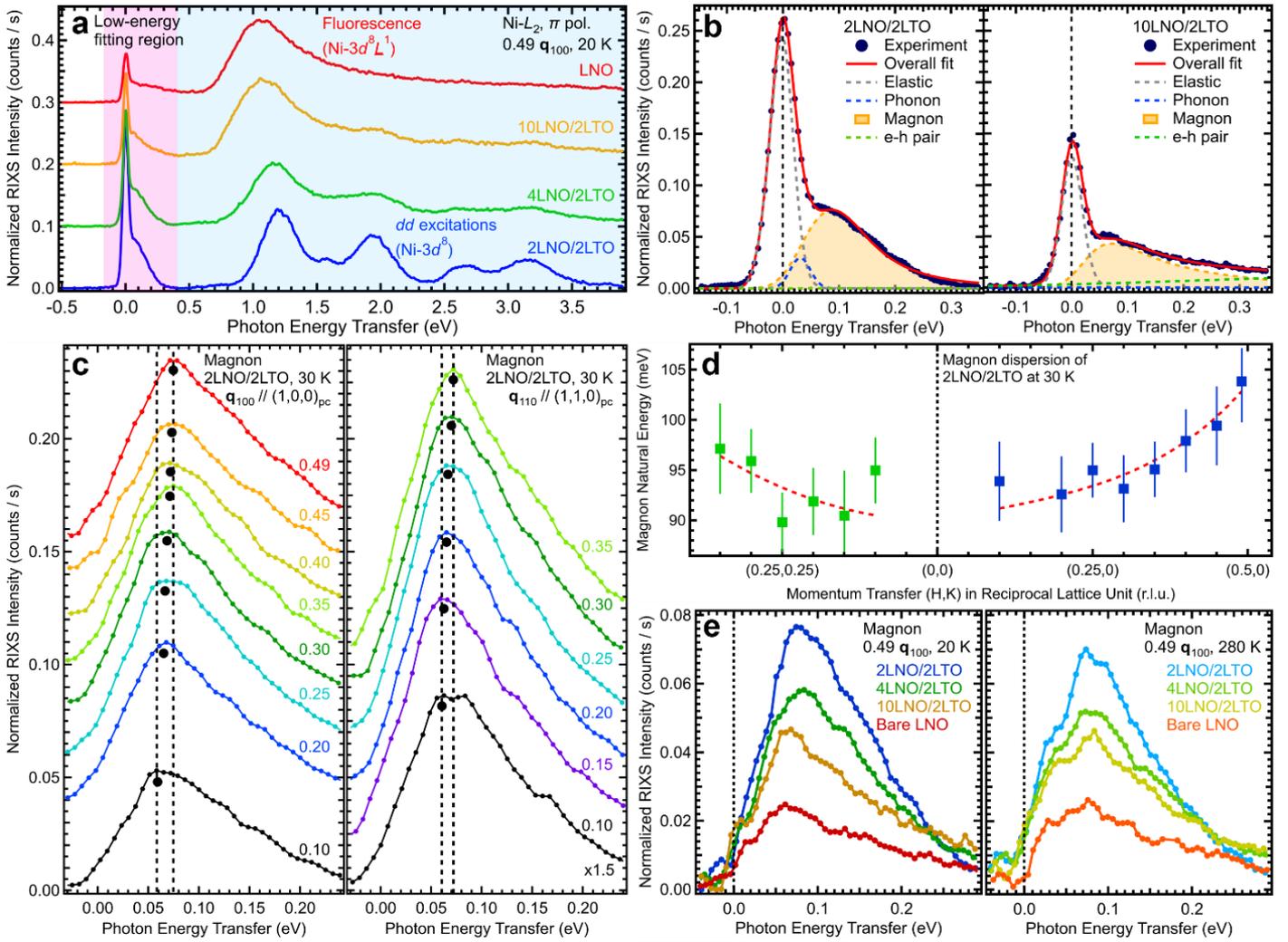

**Figure 5.** Magnetic excitations in LNO/LTO superlattices. (a) High-resolution resonant inelastic x-ray scattering (RIXS) spectra of the LNO/LTO and LNO samples, which show significant spectral features at low energy transfer. (b) Representative fitting analyses of the low-energy RIXS spectra to extract the magnon excitations, showing the cases for 2LNO/2LTO and 10LNO/2LTO. (c) Energy dispersion of the extracted magnon excitations of 2LNO/2LTO along the $(1,0,0)_{pc=pseudocubic}$ and $(1,1,0)_{pc}$ momentum directions (expressed in reciprocal lattice units of $|q_{100}| \equiv 1.60$ Å$^{-1}$ and $|q_{110}| \equiv 2.26$ Å$^{-1}$, respectively). The black dots denote the apparent center energy of the magnon, $\omega_{cen}$, at each momentum transfer, and the two striped lines denote the minimum and maximum $\omega_{cen}$ within the measurement range. As the magnon is underdamped (damping coefficient ~80 meV) and modulated by the Bose-Einstein distribution, $\omega_{cen}$ appears to be lower than the natural/undamped energy of the magnon, $\omega_{mag}$. (d) The $\omega_{mag}$ extracted from the fitting analysis of the magnon, is showing noticeable dispersion along both the $(1,0,0)_{pc}$ and $(1,1,0)_{pc}$ directions. The red striped lines are guides to the eye. The error bars denote one standard deviation. (e) Comparison of the magnon excitations of different samples at low and high temperature. All high-resolution RIXS data have been self-absorption corrected, and are taken at the Ni-$L_2$ edge with $\pi$ polarization to minimize the elastic peak.

## 3. Discussion

We can assess the type of interfacial magnetic order as follows. According to theoretical calculations, $\mu_{Ni}$ of an $S = 1$ LNO/LTO heterostructure is estimated to be ~1.7 $\mu_B$.[29] If all these moments align ferromagnetically, the total moment in LNO/LTO should approximately be $M = \mu_{Ni} n_{Ni} V_{SL} \approx 1.6 \times 10^{-4}$ emu ($n_{Ni} \approx 8.7$ nm$^{-3}$ is the number of Ni per unit volume and $V_{SL} \approx (5000 \times 5000 \times 0.046)$ μm$^3$ is the 2LNO/2LTO superlattice volume). However, when the samples are probed using magnetometry (sensitivity ~5 × 10$^{-8}$ emu), no net magnetic moment is detected. This indicates that LNO/LTO is not ferromagnetic and thus likely to be antiferromagnetic instead, similar to LNO/LaAlO$_3$.[5] Theoretical calculations also support this, as quasi-two-dimensional $G$-type antiferromagnetism with a wavevector of (½, ½, 0) is shown to have the lowest formation energy in $S = 1$ LNO/LTO.[29] Interestingly, La$_2$Ni$_3$O$_7$ is also predicted to have this $G$-type antiferromagnetism under high pressure,[33] which further highlights the similarity between LNO/LTO and this type of nickelate superconductor.

It should be noted that this antiferromagnetism emerges purely from electronic effects when interfacial LNO is transformed from paramagnetic Ni-3$d^8\underline{L}^1$ to antiferromagnetic Ni-3$d^8$ by the electron transfer. This process also removes the Ti-3$d^1$ electron from LTO along with the $S = ½$ spin necessary for its own antiferromagnetic order,[28] turning LTO into a diamagnetic SrTiO$_3$-like Ti$^{4+}$-3$d^0$ state. Therefore, the electron transfer effectively removes the magnetic moment of LTO and uses it as a seed to bolster the magnetic moment of interfacial LNO, demonstrating a high level of control on the electronic and magnetic properties of strongly-correlated oxide interfaces. This sets it apart from other magnetic LNO heterostructures such as LNO/LaMnO$_3$[6] and LNO/GdTiO$_3$,[7] where the magnetic effects in LNO are instead induced via (super)exchange interactions with the already ferromagnetic LaMnO$_3$ and GdTiO$_3$ without significantly diminishing their own magnetism.

Therefore, as the magnetism in LNO/LTO is purely interfacial and can exist without a strict geometrical confinement of the LNO layers (unlike in LNO/LaAlO$_3$[5]), a single LNO/LTO interface can essentially behave as an atomically thin strongly-correlated quasi-two-dimensional antiferromagnet, potentially making it more controllable by a large variety of external stimuli compared to three-dimensional magnets.[31] Furthermore, in LNO/LTO with thicker LNO (*e.g.*, 10LNO/2LTO), the paramagnetic metallic non-interfacial LNO is sandwiched between antiferromagnetic insulating interfacial LNO layers, making it an inherent magnetic/nonmagnetic/magnetic junction.[9,31,56,57] Thus, the interactions between the conducting charge carriers of non-interfacial LNO with the antiferromagnetic order of interfacial LNO may lead to emergent magnetoresistance effects which can be tuned by controlling the thickness of the LNO layers and hence the ratio between interfacial and non-interfacial LNO. Combined, these properties can allow the technological utilisation of LNO/LTO in advanced spintronic devices such as magnetic tunnel junctions and spin valves, especially in the emerging field of antiferromagnetic spintronics.[56,57]

Meanwhile, the singly-polarized residual holes of LNO/LTO with planar Ni-$3d_{x^2-y^2}$ + O-$2p_{xy}$ characteristics resemble the Fermi-level electronic structure of the normal state of La$_3$Ni$_2$O$_7$, which becomes superconducting under high pressure.[13,33] The quasi-two-dimensional interfacial antiferromagnetic order of LNO/LTO, with its large $\mu_{Ni}$ and $\omega_{mag}$ indicative of strong antiferromagnetic correlations, is also reminiscent of the magnetism of the parent compounds of superconducting cuprates,[10,11] (Nd,Sr)NiO$_2$,[12,53] and the predicted magnetic order of highly-pressurized La$_3$Ni$_2$O$_7$,[33] but with an $S \rightarrow 1$ spin state instead. Furthermore, the 2LNO/2LTO layered structure (two stacks of NiO$_6$ octahedra each separated by LTO layers) is also analogous to that of La$_3$Ni$_2$O$_7$ (two stacks of NiO$_6$ octahedra each separated by La$_2$O$_2$ rocksalt layers[13]). Typically, a superconducting state also requires the normal state to be metallic to provide enough carriers that can form superconducting pairs,[10,11] which is an issue for 2LNO/2LTO due to its insulating nature (Figure S2). This can be resolved with additional hole or electron doping to either enlarge the singly-polarized ligand holes or introduce back some electrons into the emptied Ti-3$d$ orbitals, respectively. External strain, for example by using different substrates, can also be introduced to further increase the orbital polarization and induce similar effects as high pressure, such as to straighten the Ni-O bond which appears to be a main characteristic of the superconducting state of La$_3$Ni$_2$O$_7$ under high pressure.[13]

We predict that this interfacial charge transfer and magnetism of LNO/LTO can be replicated in other TMO heterostructures that possess the following characteristics. First, the heterostructure pairing should be made between a charge donor and a charge acceptor (playing the role of LTO and LNO, respectively), with an appropriate energy difference to enable the charge transfer from donor to acceptor. Second, the charge transfer should result in a high-spin state of the $d$-electrons. However, it is often energetically unfavourable to directly add electrons into the $d$-bands of TMOs due to their strong on-site Coulomb repulsion. Therefore, this should be easiest to achieve if the charge acceptor also has a negative or near-zero charge-transfer energy like LNO, so that there will be enough intrinsic ligand holes that can readily recombine with the transferred electrons. This can allow the spins of the $d$-electrons to interact among themselves with minimal interference from the ligand.

## 4. Conclusion

In conclusion, here we present direct evidence of the emergence of a novel quasi-two-dimensional antiferromagnetic order with strong magnetic exchange interactions at the interfaces between strongly-correlated LaNiO$_3$ and LaTiO$_3$. Their pairing in a superlattice triggers an interlayer electron transfer from the Ti-$3d^1$ state of LaTiO$_3$ to fill the Ni-$3d^8\underline{L}^1$ ligand hole of LaNiO$_3$, resulting in an antiferromagnetic order that is inherently quasi-two-dimensionally confined at the interface. Therefore, a single LaNiO$_3$/LaTiO$_3$ interface can behave as an atomically thin strongly-correlated quasi-two-dimensional antiferromagnet, opening a path for its technological utilization in advanced antiferromagnet-based spintronic devices. This interface magnetism has a high magnetic moment and magnon energy of ~1.7 $\mu_B$ and ~100 meV, respectively, implying strong quasi-two-dimensional antiferromagnetic correlations in the LaNiO$_3$ layers. Combined with the orbitally-polarized Ni-$3d_{x^2-y^2}$ + O-$2p_{xy}$ planar ligand holes that are still left in LaNiO$_3$ after the electron transfer and its layered superlattice design, this makes the electronic, magnetic, and lattice configurations of LaNiO$_3$/LaTiO$_3$ closely resemble the precursor states of nickelate and cuprate superconductors, but with an $S \rightarrow 1$ spin state instead.

## 5. Experimental Section

*Sample Fabrication with Pulsed-laser Deposition (PLD)*

The ($n$LNO/2LTO)$_{30}$ superlattice samples ($n$ = 2, 4, 10), as well as bare LNO and bare LTO thin film samples for reference, are fabricated using the PLD system at the SIS beamline of the Swiss Light Source. LNO and LTO possess the same La *A* cation and only differ in their *B* cation (Ni and Ti, respectively), which should minimize interdiffusion between the layers. Therefore, this makes LNO/LTO an ideal system to study electronic effects, as we can be sure that any emergent phenomena only come from these electronic effects. For this purpose, we also minimize epitaxial strain to the LNO layers by using lattice-matched (110)$_{orthorhombic}$ NdGaO$_3$ single crystals ((5 × 5 × 0.5) mm$^3$ for x-ray spectroscopy experiments and (10 × 5 × 0.5) mm$^3$ for muon experiments) as the substrates. For reference, the (pseudo)cubic lattice constants of bare NdGaO$_3$, LaNiO$_3$, and LaTiO$_3$ ($a_{NGO}$, $a_{LNO}$, and $a_{LTO}$) are 3.858 Å, 3.837 Å, and 3.96 Å, respectively. The ablation targets are sintered pellets of stoichiometric LaNiO$_3$ and LaTiO$_3$ (diameter: 25 mm). The ablation is performed using an Nd:YAG pulsed laser (fourth-harmonic mode, 266 nm) with 75 mJ per pulse at a repetition rate of 10 Hz. The substrates are heated by using a CO$_2$ continuous-wave laser heater. The growth temperature is kept at 700 °C (monitored by a pyrometer) with a ramp-up rate of 8 °C per minute. Before the growth, the deposition chamber is pumped down to 2 × 10$^{-7}$ mbar to remove any excess O$_2$.

Previously, one key factor that has not been properly considered in the growth of LNO/LTO is the apparent incompatibility between the growth conditions of stoichiometric LNO (requiring high ambient growth pressure[26]) and LTO (requiring ultralow oxygen partial pressure[37]). Typically, LNO/LTO was grown in conditions favouring only LNO,[26] resulting in non-stoichiometric LaTiO$_{3+x}$ (Ti-3$d^{1-2x}$) which can transfer less than one electron per unit cell (uc) and cannot completely fill the LNO ligand hole. This is shown by the O-*K* XAS spectrum of a previous 2LNO/2LTO sample grown in these LNO-favouring conditions,[26] where the O-Ni hybridized peak (a fingerprint for the LNO ligand hole) was not quenched despite the electron transfer from LTO. To solve this issue, $N_2$ is used as the process gas during the deposition instead of the usual $O_2$ to keep an ambient growth pressure of $6×10^{-2}$ mbar. The low $O_2$ mole fraction in pure $N_2$ gas ($<10^{-5}$) results in an estimated $O_2$ activity equivalent to $<10^{-6}$ mbar suitable for the growth of stoichiometric $LaTiO_3$,[37] while the use of high $N_2$ ambient pressure of $>10^{-2}$ mbar is adequate to stabilize the growth of highly-crystalline $LaNiO_3$. This method has been shown to result in an almost complete quenching of the O-Ni hybridized peak (Figure 3b), indicating that the LNO ligand holes have almost all been filled by the electron transfer from LTO.

To fabricate the LNO/LTO superlattices in a controlled layer-by-layer manner, the $LaNiO_3$ target is ablated alternatingly with the $LaTiO_3$ target by programmable rotations of the target carousel. For each superlattice sample, the $n$LNO/2LTO supercell is repeated 30 times to ensure sufficient thickness and good signal-to-noise ratio for the x-ray and muon spectroscopy experiments. The growth is monitored *in situ* using a reflection high-energy electron diffraction (RHEED) system to ensure high crystallinity of the samples (Figure S1a – S1e). After deposition, the samples are cooled down with a rate of 16 °C/minute while maintaining the same $N_2$ ambient pressure.

*Sample Characterizations with X-ray Reflectivity (XRR) and Diffraction (XRD)*

The $LaNiO_3$ and $LaTiO_3$ layer thicknesses of each superlattice sample ($d_{LNO}$ and $d_{LTO}$) are calibrated *ex situ* using XRR measurements (Bruker D8, Figure S1f – 1h). The results show that the layer thickness can be controlled within a maximum deviation of ±13% from the ideal values based on $a_{LNO}$ and $a_{LTO}$. The interface roughness parameters ($σ_{LNO}$ and $σ_{LTO}$) are also all below 1 uc (≤ 0.3 nm). Considering that $σ_{LNO}$ and $σ_{LTO}$ are averaged over a large x-ray beam spot area of ~(2 × 0.2) $mm^2$ and thus inevitably include planar defects such as stacking faults, their small values indicate the presence of atomically flat and sharp interfaces.

The crystallinity of the samples is characterized using XRD. Despite the unorthodox use of $N_2$ as the process gas, bare $LaNiO_3$ and bare $LaTiO_3$ films show a distinct single-crystal peak with no mixture with other phases (Figure S2a). The XRD of 2LNO/2LTO (Figure S2b) shows a single main peak flanked by symmetric satellite superlattice peaks (SL-1 and SL+1), indicating its high crystallinity. Meanwhile, the main peaks of 4LNO/2LTO and 10LNO/2LTO (while also flanked by symmetric superlattice peaks) are split into double peaks.

To investigate this, the reciprocal space maps (RSM) of the samples (Figure S2c – S2e) are measured around the $(335)_{orthorhombic}$ reflection of $NdGaO_3$ (equivalent to $(103)_{pc}$). From the inverse of the reciprocal lattice vector values ($Q_x$ and $Q_z$), the average in-plane and out-of-plane pseudocubic lattice constant ($a_x$ and $a_z$) of 2LNO/2LTO can be estimated to be 3.94 Å and 3.98 Å, respectively, which is much closer to $a_{LTO}$ than $a_{LNO}$. In contrast, the average $a_x$ and $a_z$ of 10LNO/2LTO are both around 3.90 Å, *i.e.*, halfway between $a_{LNO}$ and $a_{LTO}$. These results can be explained if the LNO unit cells closer to the LNO/LTO interfaces are strongly strained by the LTO layer lattice constants. Away from the interface, the LNO unit cells start to relax towards $a_{LNO}$, bringing the average lattice constants to halfway between $a_{LTO}$ and $a_{LNO}$. Therefore, the double main peaks in the XRD of 4LNO/2LTO and 10LNO/2LTO (Figure S2b) represent the strained and relaxed parts of the LNO layers, indicating yet another dichotomy between interfacial and non-interfacial LNO, respectively. Furthermore, it is possible that this strain at the LNO/LTO interface might facilitate the interfacial charge transfer from LTO to LNO, by increasing the hybridization between the Ti, O, and Ni orbitals at the interface.[7]

The average periodicity of each superlattice can be estimated by using Bragg analysis ($n\lambda = 2\Delta \sin\theta$), where *n* is the order of the superlattice peak and *Δ* is the superlattice periodicity) on the angular distance between its SL-1 and SL+1 peaks. Based on this analysis, the average periodicity of each sample is found to be as follows. For 2LNO/2LTO it is 4.6 uc (ideal periodicity: 2LNO + 2LTO = 4 uc), for 4LNO/2LTO it is 6.7 uc (ideal periodicity: 4LNO + 2LTO = 6 uc), and for 10LNO/2LTO it is 10.3 uc (ideal periodicity: 10LNO + 2LTO = 12 uc). This means that the actual average periodicities of the superlattices can be controlled to be within ±14% of their respective ideal periodicity, consistent with XRR results.

*Sample Characterizations with Electrical Transport and Magnetometry*

DC resistivity measurements (Figure S2f) are carried out using a Physical Property Measurement System (PPMS-9T) from Quantum Design. The standard four-point technique in the van-der-Pauw configuration is used with an excitation current of 0.1 mA. The resistivity of 2LNO/2LTO is above the measurement limit, and it decreases as the thickness of LNO layers increases, with 10LNO/2LTO having a subtle insulator-to-metal transition at ~135 K. Bulk magnetometry experiments are also attempted on the LNO/LTO samples using a superconducting quantum interference device – vibrating sample magnetometer (SQUID–VSM) with a sensitivity of ~5 × $10^{-8}$ emu, but no net magnetic moment is found, indicating that the samples are not ferromagnetic and most likely antiferromagnetic instead.

*Muon Spin Rotation (µSR) Experiments*

The µSR experiments are performed by using 100% spin-polarized positive muons to interact with the local magnetic field, $B_{local}$, within the samples.[35] By tuning the muon energy into low-energy keV scale, the muon implantation depth can be controlled to be within a short distance from the sample surface, making it suitable to probe the local magnetic properties of thin films and heterostructures. The muon stopping profiles for energies optimized for each $n$LNO/2LTO sample thickness are calculated using the Monte Carlo algorithm *TRIM.SP*,[58] so that the mean depth of the muon stopping sites falls in the middle of the superlattice (Figure S3a).

The main μSR experiments are done in a transverse field geometry (TF-μSR) at the Low-Energy Muon (LEM) Facility of the Swiss Muon Source,[38] where a weak transverse magnetic field, **B**$_{TF}$ = 10 mT is applied perpendicular to the initial muon spin polarization, **S**$_\mu$ (Figure 1b). If the sample is paramagnetic, $B_{local}$ would be uniform and nearly equal to $B_{TF}$, causing the spin of the whole muon ensemble to precess with a uniform Larmor frequency of $\omega_{Larmor} = \gamma_\mu B_{local}$, where $\gamma_\mu$ is the muon gyromagnetic ratio. This manifests as an oscillation (Figure 1d) in the time-dependent spin polarization asymmetry, $A(t)$, which is modelled as

$$A(t) = a_{pm} \exp(-\Lambda t) \cos(\omega_{Larmor} t + \phi), \qquad (1)$$

where $t$ is the time after muon implantation, $a_{pm}$ is the amplitude of the oscillating component (proportional to the paramagnetic volume fraction, $f_{pm}$), $\Lambda$ is the depolarization rate, $\varphi$ is the initial phase of the muon spin with respect to the positron detector, and $\omega_{Larmor}$ is taken to be equal to $\gamma_\mu B_{TF}$. This model is used to fit the μSR data using the *MUSRFIT* program.[59] When the sample becomes magnetic, the magnetic volume fraction, $f_{mag}$, would increase, causing $f_{pm}$ and thus $a_{pm}$ to decrease. Therefore, the quenching of $A(t)$ oscillation is a direct indication of a magnetic transition in the samples, as the intrinsic moments of an ordered magnetic state would modulate **B**$_{TF}$ so that $B_{local}$ is not uniform anymore.

The sample temperature is varied using a He cryostat (temperature stability: 0.1 K) to observe the changes in $a_{pm}$. This temperature-dependent change of $a_{pm}$ is then used to extract the temperature-dependent magnetic volume fraction of the samples, $f_{mag}(T)$, and hence determine their magnetic transition temperatures (Fig. 1e), using

$$f_{mag}(T) = 1 - \frac{a_{pm}(T)}{a_{pm}(T_{max})}, \qquad (2)$$

where $a_{pm}(T_{max})$ is the amplitude in the high-temperature paramagnetic phase.

To support the TF-μSR data, we also perform temperature-dependent μSR experiments in a zero-field geometry (ZF-μSR), where no external field is applied and $A(t)$ depends solely on the intrinsic magnetic moments of the samples (Figure S3b – S3d). The ZF-μSR data can be fit by

$$A(t) = a[f \exp(-\Lambda_F t) + (1-f) \exp(-\sigma^2 t^2/2) \exp(-\Lambda_s t)], \qquad (3)$$

where $a$ is the total asymmetry, $f$ is the fraction of the fast decay part, $\Lambda_F$ and $\Lambda_S$ are the fast and slow depolarization rates, respectively, and $\sigma$ is the Gaussian contribution of the slow depolarizing part which includes the nuclear damping in the paramagnetic phase. In the high-temperature paramagnetic phase, $A(t)$ takes a long time to decay as it is governed by the slow component. As the temperature decreases below $T_{onset}$, the fraction of the fast component gradually increases and eventually dominates, leading to a quick decay of $A(t)$ at low temperatures. This bi-exponential decay is a signature of a transition into a magnetically ordered state. This result is similar to the ZF-μSR data of the LNO/LaAlO$_3$ superlattice, which also shows the quick decay of $A(t)$ below its Néel temperature.[5] The absence of a unique muon precession frequency in the ZF geometry can be explained by the presence of several inequivalent muon stopping sites within the samples: magnetic (interfacial LNO unit cells) and non-magnetic (non-interfacial LNO and LTO unit cells), which reinforces the quasi-two-dimensional nature of the magnetic order. Additionally, it is also possible that the muon precession frequency is too high to be detected by the Low-Energy μSR spectrometer.

*X-ray Absorption (XAS) and Emission (XES) Spectroscopy Experiments*

The XAS experiments (Figure 2 and 3) are done at the Ni-$L_{2,3}$, O-$K$, and Ti-$L_{2,3}$ edges using the ADRESS beamline of the Swiss Light Source at Paul Scherrer Institute.[39,40] The absorption spectra, $I$, are obtained in total fluorescence yield (TFY) mode, as total electron yield (TEY) mode is not available for the insulating 2LNO/2LTO. They are measured using both σ and π polarizations with a $\theta = 20°$ incident angle from the sample surface. The in-plane, $I_{xy}$, and out-of-plane, $I_z$, absorption spectra are obtained from $I_\sigma$ and $I_\pi$ using the Malus' law according to

$$I_{xy} = I_\sigma \tag{4}$$

$$I_z = [I_\pi - I_{xy} \sin^2\theta]/\cos^2\theta. \tag{5}$$

The TFY spectra are normalized and self-absorption corrected using the *normalization* and *self-absorption* algorithms of the *ATHENA* XAS data processing program.[60] Unfortunately, the Ni-$L_3$ spectra are dominated by strong La-$M_4$ absorption (Figure S4). Therefore, when discussing the differences between the Ni-$L_{2,3}$ spectra of the samples in Figure 2a, we focus on the Ni-$L_2$ edge as Ni-$L_2$ is free from this La-$M_4$ contamination. The peaks are fitted with pseudo-Voigt functions, while the step-like background (coming from the excitation from the Ni-$2p$ core states into the continuum) is fitted with Gaussian error functions (erf). In Figure 2a, this background is removed after applying self-absorption correction to allow better comparison between the experimental and the calculated XAS spectra.

The O-$K$ XES spectra are obtained by taking off-resonant inelastic x-ray scattering at 5.5 eV above the O-Ni resonance (Figure 3b) using the ADRESS beamline.[39,40] They are measured using both σ and π polarizations with $\theta = 20°$ incident and $\alpha = 130°$ scattering angles and converted into the in-plane and out-of-plane components using Equation 4 and 5, respectively. As the emission energies of the O-$2p$ occupied DOS are far below any of the O-$K$ absorption peaks, the XES spectra are not affected by self-absorption effects. In all XAS and XES experiments, the sample temperature is kept at 25 K using a He cryostat with a temperature stability of 1 K. The Ti-$L_3$ and O-$K$ XAS and XES spectra of the samples along the out-of-plane **E** // *z* and in-plane **E** // *xy* polarization are shown in Figure 3 and Figure S5, respectively.

*Resonant Inelastic X-ray Scattering (RIXS) Experiments*

The medium-resolution (FWHM *ca.* 110 meV) RIXS energy maps (Figure 2b – 2e) are taken at the Ni-$L_3$ edge using the ADRESS beamline[39,40] with σ polarization, $\theta$ = 20° incident angle, $\alpha$ = 130° scattering angle, and 25 K sample temperature (Figure 7a). As the emission energy of the signature charge-transfer (CT) excitation is far below any of the Ni-$L_3$ absorption peaks, it is not affected by self-absorption effects.

The high-resolution RIXS experiments (Figure 5) are done using the ID32 beamline of the European Synchrotron Radiation Facility[50] (for momentum-dependent studies) and the I21 beamline of the Diamond Light Source[51] (for sample- and temperature-dependent studies). The spectra are taken at the Ni-$L_2$ resonance with π polarization to reduce the elastic scattering contributions, especially from the La-$M_4$ contamination near the Ni-$L_3$ edge (Figure S4). To reach certain photon momentum transfer, **q**, in a geometry along the in-plane $(1,0,0)_{pc}$ and $(1,1,0)_{pc}$ momentum directions (expressed in reciprocal lattice units (r.l.u.) of $|\mathbf{q}_{100}| \equiv 1.60$ Å$^{-1}$ and $|\mathbf{q}_{110}| \equiv 2.26$ Å$^{-1}$, respectively) while minimizing the elastic line and the out-of-plane **q** component, both $\theta$ and $\alpha$ are rotated in tandem according to Table S1, so that $\theta$ is kept in near-grazing geometry. For the **q**-dependent study at ID32, the temperature is kept at 30 K, while for the temperature-dependent study at I21, the measurements are done at 20 K and 280 K. All RIXS experiments are done using He cryostats with a temperature stability of 1 K, and the detected scattered photons are not filtered by the outgoing polarization. Due to the different specifications of each beamline, the ID32 data are taken with the single-photon counting mode, while the I21 data are taken with the integration mode. For consistency, the I21 data are normalized to the ID32 data.

As the emission energies of the low-energy features in Figure 5 are still very close to the incident photon energy at the maximum of the Ni-$L_2$ absorption edge, the intensity of the high-resolution RIXS spectra is reduced by self-absorption effects by a factor of

$$C(\omega_1, \omega_2, \theta, \alpha) = \frac{1}{1 + \frac{\mu_2(\omega_2) \sin\theta}{\mu_1(\omega_1)\sin(\alpha-\theta)}}, \quad (6)$$

where $\mu_1$ ($\mu_2$) is the normalized absorption coefficient of the incident (outgoing) photon at a particular incident (outgoing) energy, $\omega_1$ ($\omega_2$), as shown in Figure 2a. The self-absorption-corrected (SAC) RIXS data are obtained by dividing the raw RIXS spectra by *C*. As the low-energy features of the high-resolution RIXS spectra are dominated by spin-flipping magnetic excitations and the incident polarization is π, the self-absorption corrections are performed under the assumption that the outgoing photons are mostly σ-polarized, *i.e.*, in the spin-flip πσ' scattering channel (Figure S6 – S8).

The SAC Ni-$L_2$ high-resolution RIXS spectra are fitted using three fitting components (Figure 5b, Figure S6 – S8): a Gaussian for the elastic line, a Gaussian for a low-energy phonon at ~30 meV (presumably an octahedral-distortion phonon[61]), and a damped harmonic oscillator[51] according to

$$I_{\mathrm{mag}}(\omega) = A_{\mathrm{mag}} n_{BE} \frac{4\sigma_{\mathrm{mag}}\omega_{\mathrm{mag}}\omega}{\left(\omega^2-\omega_{\mathrm{mag}}^2\right)^2+\left(2\omega\sigma_{\mathrm{mag}}\right)^2} \quad (7)$$

for the damped magnon excitation. Here, $A_{\mathrm{mag}}$ is the amplitude of the magnon, $\sigma_{\mathrm{mag}}$ is the magnon damping coefficient, $\omega_{\mathrm{mag}}$ is the natural/undamped magnon energy (Figure 5e), and $n_{BE}$ is the Bose-Einstein (BE) temperature factor according to

$$n_{\mathrm{BE}}(\omega, T) = \frac{1}{1-\exp\left(-\frac{\hbar\omega}{k_{\mathrm{B}}T}\right)}, \quad (8)$$

where $T$ is the sample temperature, $\hbar$ is the reduced Planck constant, and $k_B$ is the Boltzmann constant. For more metallic samples, an additional Gaussian is also added to represent the electron-hole (e-h) pair excitations at higher energy transfers (> 0.3 eV). The extracted magnon excitation in Figure 5c – 5e is then obtained by subtracting the elastic, phonon, and e-h pair spectral weights from the SAC spectra.

Unfortunately, due to technical limitations of the beamlines, the high-resolution RIXS data is measured only at temperatures above 20 K, *i.e.*, above the $T_{\mathrm{mag}}$ of 18 K, resulting in a broadened and damped magnon. This also prevents us from probing the microscopic structure of the long-range magnetic order, for example by using the resonant x-ray diffraction technique. Therefore, detailed theoretical analysis of the magnon dispersion, for example to extract the quantitative value of the magnetic exchange coupling, is not feasible within the scope of this study.

*Cross-Sectional Scanning Transmission Electron Microscopy (STEM) and Electron Energy Loss Spectroscopy (EELS) Experiments*

To investigate its detailed atomic structure and spatially resolve its electronic state, we performed cross-sectional STEM imaging and EELS on the 10LNO/2LTO sample. The TEM specimens are prepared using a focused Ga-ion beam tool (FIB: Zeiss, Nvision 40) to cut a lamella of ~50 – 100 nm thickness from the sample material. The sample is first etched using a 30 keV Ga-beam (80 pA and 40 pA beam current), then it is polished with a 5 keV Ga-beam to reduce the thickness of the amorphized surface layer. During this process, the sample surface is protected by a thick Pt layer.

The STEM experiments are performed on a Cs probe-corrected FEI Titan³ G2 60 – 300 microscope (Thermo Fisher Scientific, Eindhoven) equipped with a field emission source (X-FEG), a GIF Quantum ERS from Gatan for EELS including a K2 Summit direct electron detection camera and a monochromator. The microscope is operated at 300 kV in scanning mode. The beam current is set to ~100 pA with convergence semi-angles of 19.6 mrad and 15.5 mrad (for monochromated EELS). The angular range for high-angle annular dark-field (HAADF) imaging is 62.2 – 214.0 mrad. For EELS the collection semi-angle is 13 mrad and the microscope is operated in monochromatized mode with an energy resolution of ~0.2 eV (measured by the FWHM of the zero-loss peak). All data is taken from an area with a relative thickness of 0.4 – 0.6 in units of inelastic mean free path, which corresponds to a sample thickness of ~50 – 76 nm at the given conditions[62]. STEM imaging, EELS acquisition, and analysis are performed using GMS 3 (version 3.60) by Gatan. For EELS analysis the background is removed using a power-law pre-edge fit and the signal-to-noise ratio is improved using the first eight components of a principle component analysis (PCA).

The HAADF STEM image of 10LNO/2LTO (Figure S9a) shows a clear distinction between the LNO and LTO layers with minimal interdiffusion between the layers. Some local dislocations (orange arrows in Figure S9a) and crystal domains do exist at the interface, which are presumably caused by the strain stemming from the sizeable difference between the lattice constants of LNO and LTO ($a_{LNO}$ = 3.837 Å vs $a_{LTO}$ = 3.96 Å). Nevertheless, the cross-sectional O-$K$ EELS spectra (Figure S9b) show that there are clear differences between the electronic states of interfacial and non-interfacial LNO. In particular, as the charge transfer and thus the modification of the interfacial electronic state of LNO/LTO are identified by the suppression of the O-Ni peak (Figure 3b), we performed a fitting analysis of the EELS spectra (Figure S9c) to disentangle it from other, much stronger peaks and spatially resolve it across the interface. The results (Figure S9d) show that the O-Ni peak of interfacial LNO is very much reduced, indicating a high-spin Ni-$3d^8$-like state at the interface due to the charge transfer from LTO. In contrast, the O-Ni peak at the LNO unit cells away from the interface remain strong, signifying that their electronic state remains unchanged from their bare-LNO-like Ni-$3d^8\underline{L}^1$ state. Therefore, this is a direct verification of the analysis in Figure 4, which states that the electron transfer from LTO (and thus the magnetism) is confined only to the interface. Meanwhile, Figure S9e shows that the interfacial LNO unit cells also have a considerable O-Ti spectral weight, which is not surprising as some of their oxygen ions are shared with LTO.

*Theoretical Calculations*

The theoretical calculations for the ground states and spectral properties of LNO/LTO superlattices are performed using the *Quanty* package[41,42] based on a single impurity Anderson model (SIAM). Double cluster models have been successful in describing nickelates exhibiting breathing distortions and associated bond disproportionation,[18] but given the absence of this disproportionation in LNO, the SIAM is a more suitable model as both the charge transfer continuum and fluorescence excitations can be appropriately captured. The general form of our SIAM Hamiltonian, $H$, is

$$H = H_I + H_L + H_V + H_{\text{res}}, \tag{9}$$

where $H_I$ and $H_L$ are the impurity and ligand (bath) Hamiltonians, respectively, while $H_V$ is their hybridization interaction. The $H_{res}$ is an electron doping reservoir weakly coupled to the system, similar to that used in the double cluster approach of electron-doped $R$NiO$_3$.[43]

The $H_I$ is a conventional multiplet crystal field Hamiltonian which includes multipole Coulomb interactions, octahedral crystal field potentials, as well as spin-orbit coupling in the 2$p$ core and 3$d$ valence shells (see our previous study for a full description[18]). The $H_I$ has the form

$$H_I = H_U^{dd} + H_U^{pd} + H_{l\cdot s}^d + H_{l\cdot s}^p + H_o^d + H_o^p \tag{10}$$

where $H_U^{dd}$ and $H_U^{pd}$ are the 3$d$-3$d$ and 2$p$-3$d$ Coulomb repulsion (including all multiplet effects), respectively, while $H_{l\cdot s}^d$ and $H_{l\cdot s}^p$ are the 3$d$ and 2$p$ spin-orbit interactions, respectively. Lastly, $H_o^d$ and $H_o^p$ are the 3$d$ and 2$p$ electron on-site energies, respectively, which (for the 3$d$ shell) also include the crystal fields. Parameters which enter into this portion of the Hamiltonian include the monopole Coulomb interactions ($U_{dd}$ and $U_{pd}$), the radial Slater integrals ($F_{ii}^k$ and $G_{ij}^k$), the spin-orbit interaction energies ($\zeta_i$), and the crystal field energy ($10Dq$), as given in Table S2. These parameter values are similar to those used previously in our bare $R$NiO$_3$ study.[18] The Hamiltonian also includes the exchange interaction energy of 120 meV, based on the value for NiO which has a similar antiferromagnetic Ni-3$d^8$ state.[30] For $10Dq$, we provide separate values for the undoped (Ni-3$d^8\underline{L}^1$) and fully-doped (Ni-3$d^8$) cases (Table S2), and we linearly interpolate between these values for intermediate doping. The primary $dd$ excitation of the RIXS spectra is sensitive to the $10Dq$ value, which enables this doping-dependent evaluation.

The ligand bath term of the SIAM Hamiltonian has the form

$$H_L = \sum_{i=1}^{N_V} \varepsilon_L L_i^\dagger L_i + \frac{W}{4} \sum_{i=1}^{N_V-1}\left(L_i^\dagger L_{i+1} + L_{i+1}^\dagger L_i\right) + \varepsilon_C C_i^\dagger C_i, \tag{11}$$

where $L_i^\dagger$ ($L_i$) creates (annihilates) an electron in the ligand bath shell with onsite energy $\varepsilon_L$ indexed by $i$. The ligand sites comprising the bath are coupled in a chain via the second summation, leading to a bath of width $W$. While the ligand bath shells denoted by operators $L$ are fully occupied (before hybridization), those denoted by operators $C$ are unoccupied, and allow for fluctuations away from the impurity. In other words, $L_i$ ($C_i$) correspond to valence (conduction) baths. These valence and conduction baths are coupled to the impurity with the hybridization term

$$H_V = \sum_\tau \left[V_{L,\tau}\left(d_\tau^\dagger L_{1,\tau} + L_{1,\tau}^\dagger d_\tau\right) + V_{C,\tau}\left(d_\tau^\dagger C_\tau + C_\tau^\dagger d_\tau\right)\right], \tag{12}$$

where $d_\tau^\dagger$ ($d_\tau$) creates (annihilates) an electron in the Ni $d$ shell with combined orbital and spin index $\tau$. The symmetry-dependent hopping integrals $V_{L,\tau}$ and $V_{C,\tau}$ define the strength of the interaction between the impurity and valence and conduction baths, respectively. Finally, the doping term has the form

$$H_{res} = \varepsilon_R R^\dagger R + \sum_\tau V_R\left(R_\tau^\dagger L_{1,\tau} + L_{1,\tau}^\dagger R_\tau\right), \tag{13}$$

where $R_\tau^\dagger$ ($R_\tau$) creates (annihilates) an electron in a weakly coupled electron doping reservoir.[43] The level of doping is controlled by the onsite energy $\varepsilon_R$ and the weak coupling is induced by a hopping integral $V_R$. The parameters for the hybridization portion of the Hamiltonian are given in Table S3. Here, the onsite energies $\varepsilon_L$, $\varepsilon_C$, and $\varepsilon_R$ are converted to their respective charge-transfer energy representations of $\Delta_L$, $\Delta_C$, and $\Delta_R$. The hopping integrals and valence ligand charge-transfer energies shown in Table S3 are similar to those used for bond-disproportionated bare $R$NiO$_3$.[18,43] XAS and RIXS spectra are simulated using built-in *Quanty* functions, which are then broadened and normalized at the off-resonant low- and high-energy tails for comparison to experiment.

As each cluster is composed of one 2$p$ shell, one 3$d$ shell, 20 valence bath sites (each implemented as a $d$ shell), one conduction bath (implemented as a $d$ shell), and the charge reservoir (implemented as an $s$ shell), the model is composed of 228 Fermionic (spin) orbitals, having a filling of 215 electrons. The remaining 13 holes come from the 3$d$ shell (3 holes due to the 3$d^7$ nominal configuration of LaNiO$_3$) and the conduction bath shell (which, as it is simulated as an empty $d$ shell, has 10 holes). This Hilbert space has a dimension of ~5 × 10$^{20}$, which is prohibitively large. We circumvent this by restricting the Hilbert space only to configurations having at most two valence bath holes and one conduction bath hole, so that accurate spectra can still be achieved with the iterative diagonalization and response function methods employed by *Quanty*.

*Statistical Analysis*

The error bars of the magnetic volume fraction ($f_{mag}$) in Figure 1e are obtained by fitting the time-dependent TF-μSR data with Equation 1 using the *MUSRFIT* program[59] (as representatively shown in Figure 1d) to first get the amplitude of the oscillating component ($a_{pm}$) and its corresponding fitting error, which are then propagated to $f_{mag}$ using Equation 2. Meanwhile, the error bars of the magnon natural energy ($\omega_{mag}$) in Figure 5d are obtained from fitting the self-absorption-corrected high-resolution RIXS data with Equation 7 using the *lmfit* package of *Python*, as representatively shown in Figure 5b. In both cases, the error bars represent one standard deviation away from the fitted parameters.

**Acknowledgements**

The μSR experiments have been performed at the the Low-Energy Muon Facility of the Swiss Muon Source at Paul Scherrer Institute. The XAS and RIXS experiments have been performed at the ADRESS beamline of the Swiss Light Source at Paul Scherrer Institute, the ID32 beamline of the European Synchrotron Radiation Facility, and the I21 beamline of the Diamond Light Source under the session MM28665. Part of the resistivity measurements were performed at the Laboratory for Multiscale Materials Experiments of Paul Scherrer Institute. The experimental work at Paul Scherrer Institute is supported by the Swiss National Science Foundation (SNSF) through projects no. 200021_178867 and 206021_139082, the Sinergia network Mott Physics Beyond the Heisenberg Model (MPBH) (SNSF Research Grants CRSII2 160765/1 and CRSII2 141962), and the Cross-RIXS funding from Paul Scherrer Institute. T.C.A., N.K, and Y.W. acknowledge funding from the European Union's Horizon 2020 research and innovation programme under the Marie Skłodowska-Curie grant agreement No. 701647 and 884104 (PSI-FELLOW-II-3i and PSI-FELLOW-III-3i program).

**Conflict of Interest**

The authors declare no conflict of interest.

**Author Contributions**

T.C.A., M.R., and T.S. planned the project. T.C.A. and M.R. fabricated the samples and characterized them using x-ray diffraction. Y.M.K., N.K., M.M., Y.S., and T.C.A. characterized the samples using electrical transport and magnetometry. D.K., E.M., and M.R. characterized the samples using transmission electron microscopy techniques. A.S., Z.S., T.P., T.C.A., Y.W., and T.S. performed muon spin rotation/relaxation experiments. T.C.A., Y.W., W.Z., Y.T., T.Y., D.B., M.G.-F., S.A., C.W.G., K.J.Z., N.B.B., and T.S. performed x-ray spectroscopy experiments. R.G., G.H., and T.C.A. performed theoretical calculations. T.C.A. and T.S. analyzed the data. T.C.A. and T.S. wrote the manuscript with inputs from all co-authors. T.S. led the project.

**Data Availability Statement**

The data that support the findings of this study are openly available in Zenodo at https://doi.org/10.5281/zenodo.12745079.

**Supporting Information**

**Table S1.** Geometry configurations for high-resolution resonant inelastic x-ray scattering (RIXS) experiments. The incident ($\theta$) and scattering ($\alpha$) angles are set such that the momentum transfer, **q**, is mostly along the in-plane $(1,0,0)_{pc=pseudocubic}$ and $(1,1,0)_{pc}$ momentum directions (expressed in reciprocal lattice units (r.l.u.) of $|\mathbf{q}_{100}| \equiv 1.60$ Å$^{-1}$ and $|\mathbf{q}_{110}| \equiv 2.26$ Å$^{-1}$, respectively). The $\theta$ is measured from the sample surface, while $\alpha$ is measured from the incident momentum vector. The experiments are performed at the Ni-$L_2$ edge (870.7 eV).

| $\mathbf{q}_{100}$ (h, k) | $\theta$ (°) | $\alpha$ (°) | $\mathbf{q}_{110}$ (h, k) | $\theta$ (°) | $\alpha$ (°) |
|---|---|---|---|---|---|
| (0.10, 0) | 30.2 | 90.00 | (0.10, 0.10) | 23.8 | 90.00 |
| (0.20, 0) | 14.3 | 90.00 | (0.15, 0.15) | 12.2 | 90.00 |
| (0.25, 0) | 5.30 | 90.00 | (0.20, 0.20) | 8.10 | 100.0 |
| (0.30, 0) | 5.00 | 100.0 | (0.25, 0.25) | 4.70 | 111.0 |
| (0.35, 0) | 5.40 | 111.0 | (0.30, 0.30) | 5.50 | 128.0 |
| (0.40, 0) | 10.5 | 128.0 | (0.35, 0.35) | 6.70 | 149.5 |
| (0.45, 0) | 17.3 | 149.5 | | | |
| (0.49, 0) | 8.10 | 149.5 | | | |

**Table S2.** Parameters for the multiplet crystal field theory part of the Hamiltonian used in the single impurity Anderson model (SIAM) calculation. All values are in units of electron volts. The $U_{dd}$ and $U_{pd}$ are the monopole Coulomb interactions, $F^k_{ii}$ and $G^k_{ij}$ are the radial Slater integrals, $\zeta_i$ are the spin orbit interaction energies, and $10Dq$ is the crystal field energy. The index 2p and 3d correspond to the Ni 2p and 3d shells, respectively, while dd and pd indicate interactions within the metal 3d shell and between the metal 2p and 3d shells, respectively. Where two values are given, the first (second) corresponds to the undoped (fully-doped) case, and linear interpolation is used for intermediate doping.

| Configuration | $U_{dd}$ | $U_{pd}$ | $F^2_{dd}$ | $F^4_{dd}$ | $F^2_{pd}$ | $G^1_{pd}$ | $G^3_{pd}$ | $\zeta_{2p}$ | $\zeta_{3d}$ | $10Dq$ |
|---|---|---|---|---|---|---|---|---|---|---|
| No core hole | 6.00 | - | 9.29 | 5.81 | - | - | - | - | 0.091 | 0.75, 0.55 |
| With core hole | 6.00 | 7.00 | 10.62 | 6.64 | 6.68 | 5.06 | 2.88 | 11.51 | 0.091 | 0.75, 0.55 |

**Table S3.** Parameters for the bath and hybridization part of the Hamiltonian used in the single impurity Anderson model (SIAM) calculation. All values are in units of electron volts, except $N$ which is the number of bath sites. The $W$ is the ligand bath width, $\Delta_L$, $\Delta_C$, and $\Delta_R$ are the charge-transfer energies, and $V_{L,e_g}$, $V_{L,t_{2g}}$, $V_{C,e_g}$, $V_{C,t_{2g}}$, and $V_R$ are the hopping integrals. The indices $L$ and $C$ correspond to the occupied and unoccupied ligand bath sites (before hybridization), respectively, while $R$ corresponds to the electron doping reservoir. Lastly, the indices $e_g$ and $t_{2g}$ correspond to the $e_g$ and $t_{2g}$ orbitals of the metal $d$ band, respectively. Where two values are given, the first (second) corresponds to the undoped (fully-doped) case, and linear interpolation is used for intermediate doping. The $V_{L,e_g}$ and $V_{L,t_{2g}}$ are rescaled by 0.85 for the configuration with the core hole, as typically done for cluster-based simulations of spectroscopy results.

| $N$ | $W$ | $\Delta_L$ | $\Delta_C$ | $\Delta_R$ | $V_{L,e_g}$ | $V_{L,t_{2g}}$ | $V_{C,e_g}$ | $V_{C,t_{2g}}$ | $V_R$ |
|---|---|---|---|---|---|---|---|---|---|
| 20 | 7.00 | 0.00 | 8.00 | 2.00, 3.07 | 2.60, 2.20 | 1.30, 1.10 | 0.90, 0.00 | 0.45, 0.00 | 0.25 |

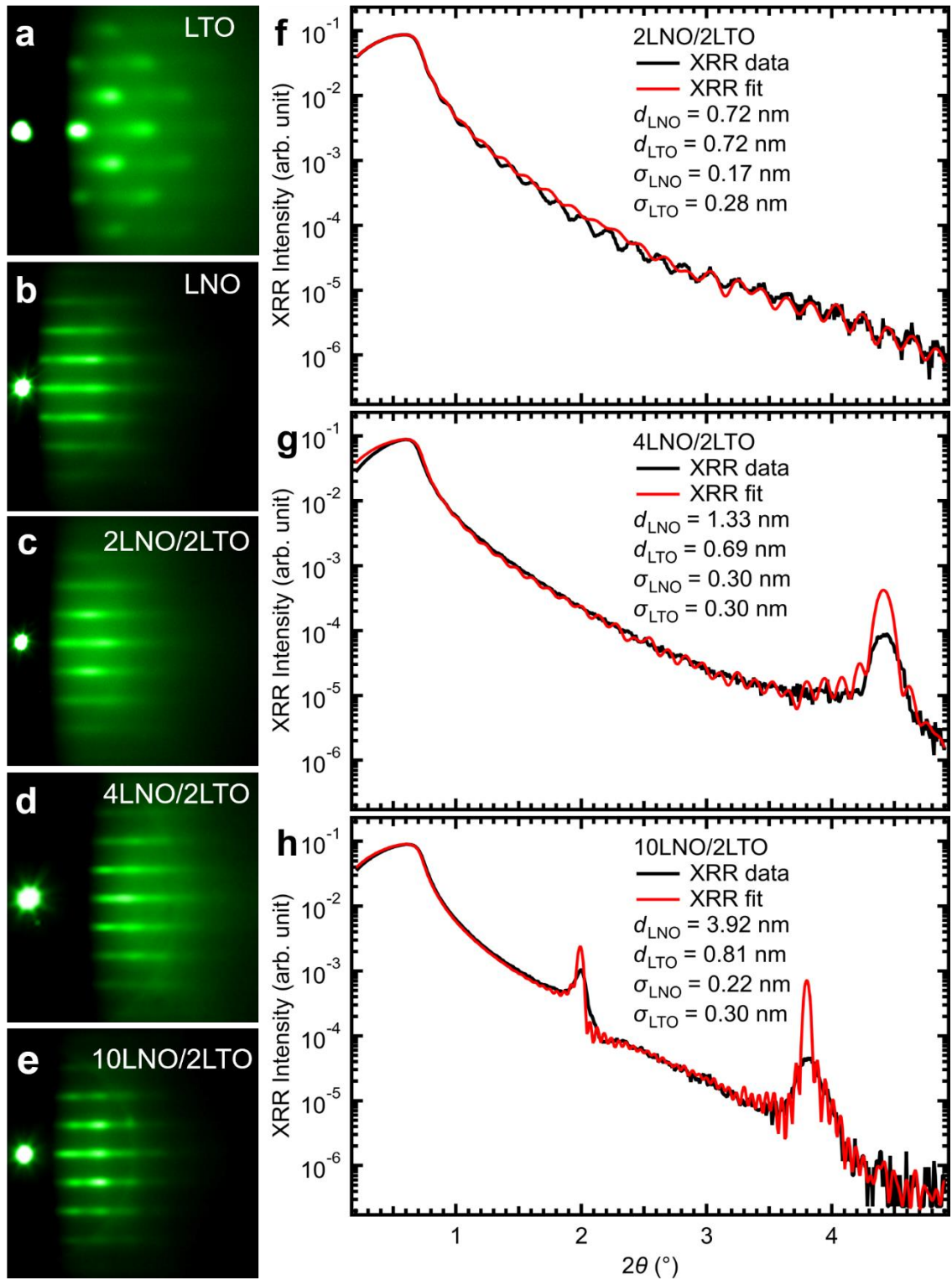

**Figure S1.** Reflection high-energy electron diffraction (RHEED) and x-ray reflectivity (XRR) of LNO/LTO superlattices. (a) – (e) RHEED images after the growth of each sample using pulsed-laser deposition, showing the high-crystallinity and two-dimensionality of the sample growths. (f) – (h) XRR of the LNO/LTO samples, showing the average thicknesses of the LNO ($d_{LNO}$) and LTO ($d_{LTO}$) layers of each sample as well as their corresponding average interface roughnesses ($\sigma_{LNO}$ and $\sigma_{LTO}$).

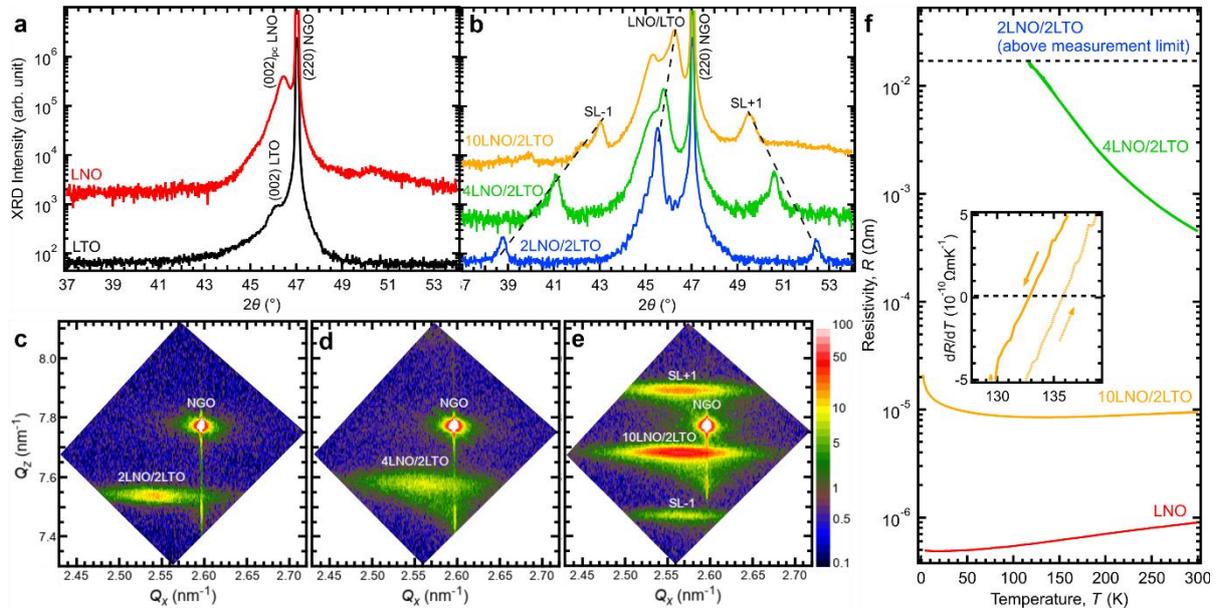

**Figure S2.** X-ray diffraction (XRD) and electrical transport of LNO/LTO superlattices. (a) and (b), XRD results of LNO/LTO superlattices, bare LNO (34 nm thick), and bare LTO (36 nm thick) samples grown on (110)$_{or=orthorhombic}$ NdGaO$_3$ (NGO) substrates. The samples all show distinct main (002)$_{pc=pseudocubic}$ peaks next to the (110)$_{or}$ NGO substrate peak. The (002)$_{pc}$ peaks of 4LNO/2LTO and 10LNO/2LTO are split into two because the LNO layers close and away from the interface experience different amount of strain by the LTO layers. The superlattice samples also show clear superlattice peaks (SL-1 and SL+1) that symmetrically flank the main peaks. (c) – (e) Reciprocal space maps of the LNO/LTO samples, taken around the (335)$_{or}$ (equivalent to (103)$_{pc}$) reflection of NGO. The $Q_x$ and $Q_z$ are the in-plane and out-of-plane reciprocal lattice vector values, respectively. In all samples, $Q_x$ and $Q_z$ of LNO layers are smaller than NGO, indicating larger lattice constants compared to NGO and therefore bulk-like LNO (as the lattice constants of NGO and bulk-like LNO are nearly the same). (f) Electrical transport characterizations of LNO/LTO superlattices. The resistivity of 2LNO/LTO is above the measurement limit. Inset shows the resistivity derivative of 10LNO/2LTO, highlighting the subtle metal-insulator transition at ~135 K.

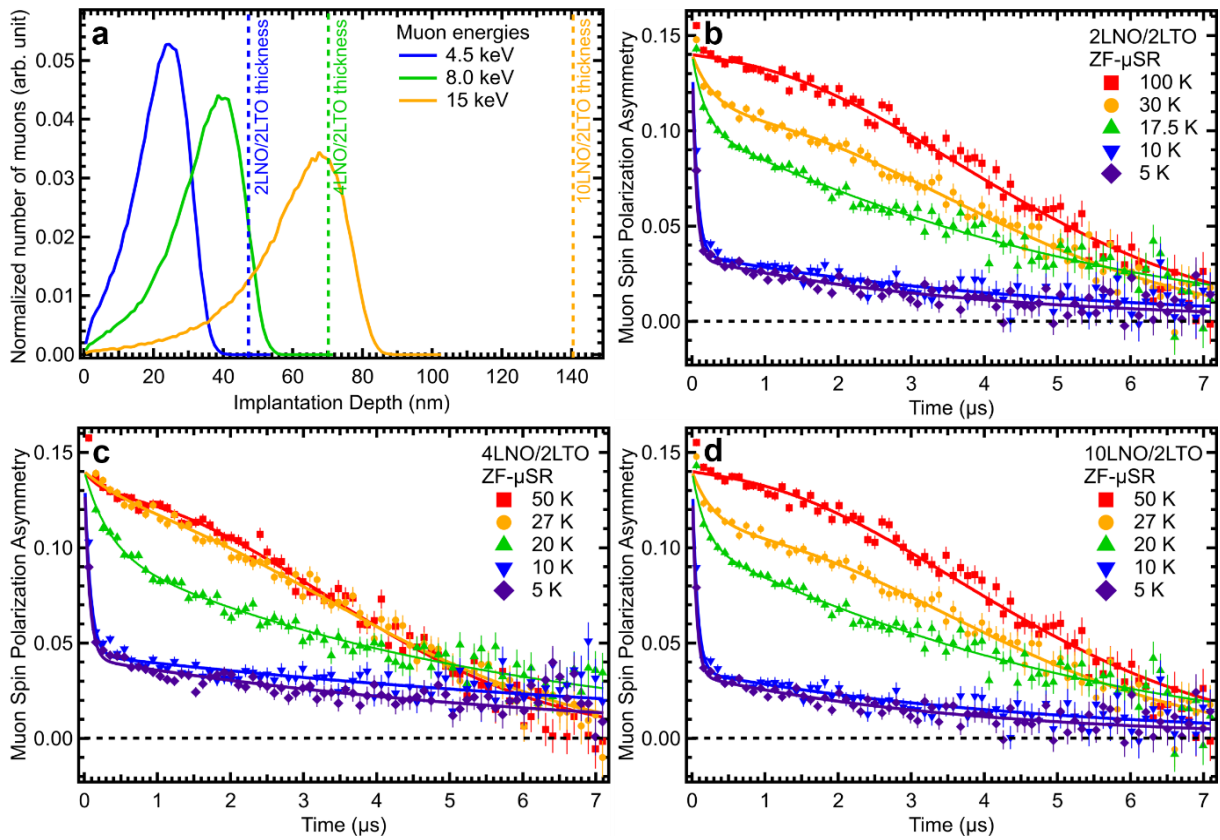

**Figure S3.** Muon spin rotation/relaxation (µSR) experiments of LNO/LTO superlattices. (a) Calculated muon stopping profiles for energies optimized for each LNO/LTO sample thickness (denoted by the dashed lines). (b) – (d) Temperature-dependent muon spin polarization asymmetry decay of the LNO/LTO samples in zero-field (ZF) µSR geometry.

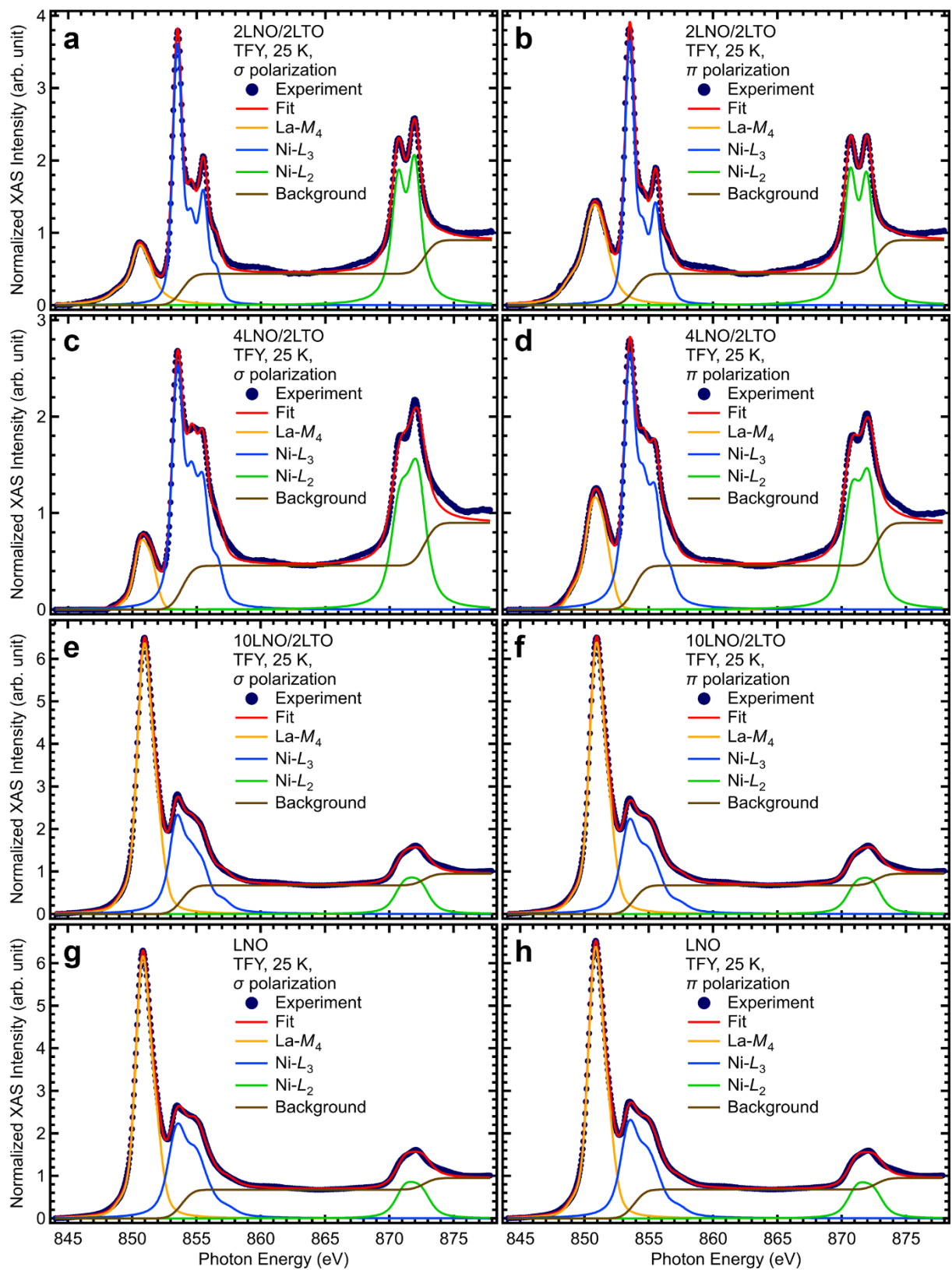

**Figure S4.** Fitting analyses of Ni-$L_{2,3}$ x-ray absorption spectroscopy (XAS) of the LNO/LTO superlattice and bare LNO samples. The spectra are fitted with pseudo-Voigt functions, while the step-like background (coming from the excitation from the Ni-$2p$ core states into the continuum) is fitted by Gaussian error functions (erf).

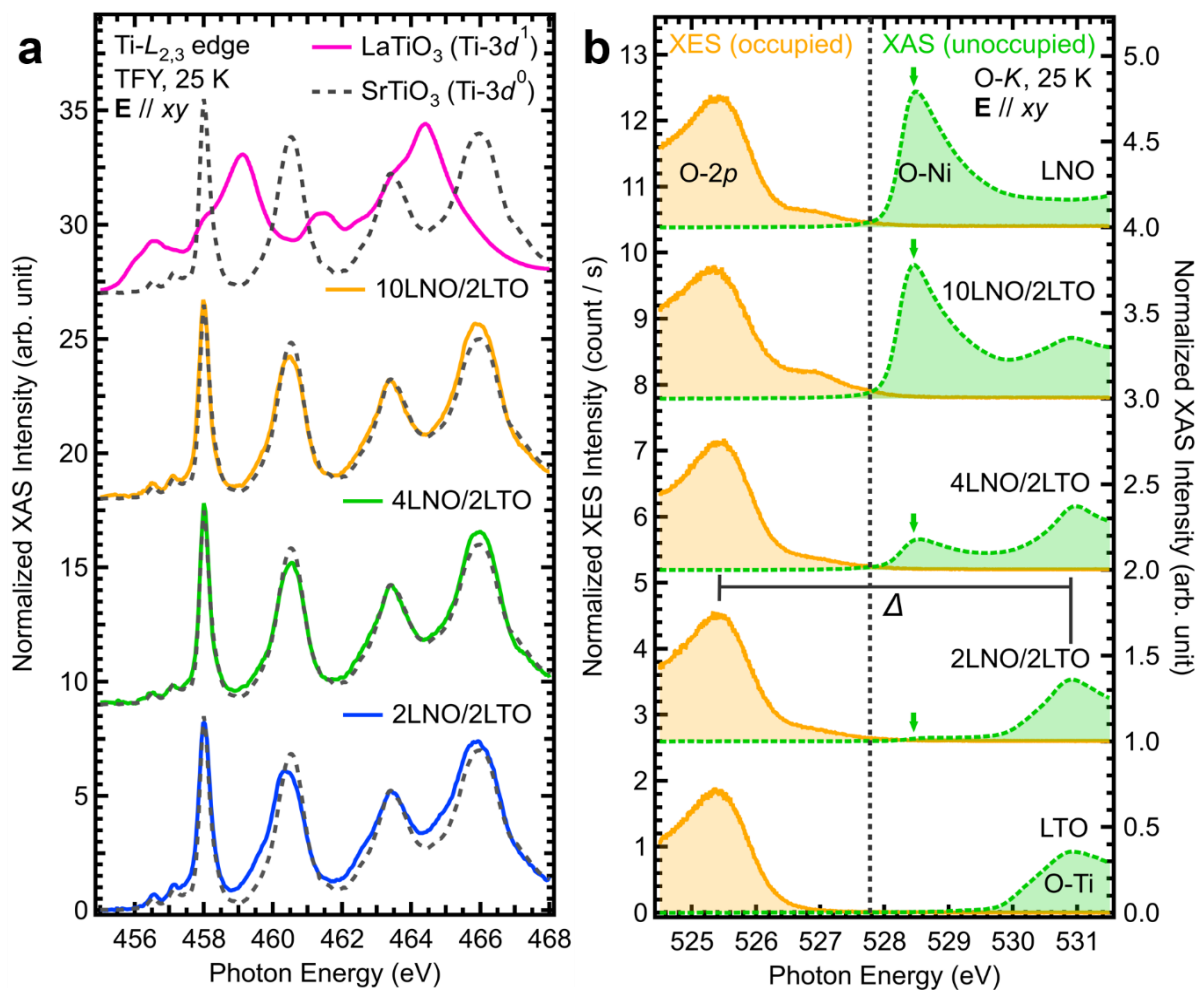

**Figure S5.** Ti-$L_{2,3}$ and O-$K$ x-ray absorption spectroscopy (XAS) of LNO/LTO superlattices for in-plane (E // *xy*) polarization. (a) Experimental Ti-$L_{2,3}$ XAS spectra of the samples. They resemble the diamagnetic Ti$^{4+}$-3$d^0$ configuration of SrTiO$_3$ instead of the antiferromagnetic Ti$^{3+}$-3$d^1$ configuration of bare LTO, signifying that the Ti-3$d$ orbitals of LTO in LNO/LTO have indeed been emptied by the electron transfer into the LNO layers. The Ti-$L_{2,3}$ XAS spectrum of bare LTO is reprinted with permission from M. Haverkort *et al.*, *Phys. Rev. Lett.* **94**, 056401 (2005)[28]. Copyright (2005) by the American Physical Society. DOI: https://doi.org/10.1103/PhysRevLett.94.056401. (b) Experimental O-$K$ x-ray emission spectroscopy (XES) and XAS of the samples (equivalent to occupied and unoccupied density of states (DOS), respectively), showing the diminishing of the O-Ni hybridized peak until a positive charge-transfer gap of $\Delta \approx 5$ eV opens for 2LNO/2LTO. The XAS data are taken with the total fluorescence yield (TFY) mode and have been self-absorption corrected, while the O-$K$ XES data are obtained by taking non-resonant inelastic x-ray scattering spectra at ~5.5 eV above the O-Ni peak.

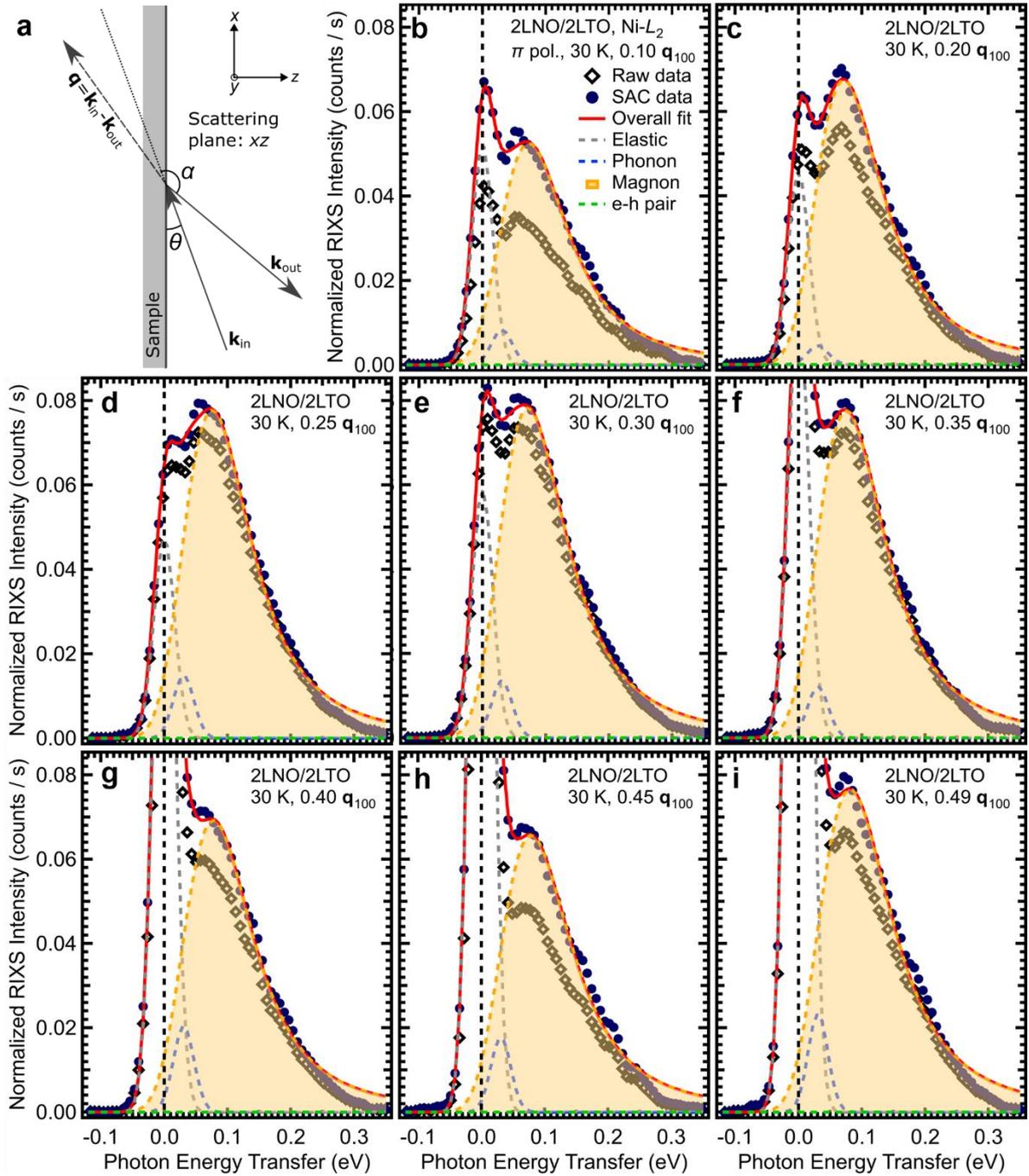

**Figure S6.** Fitting analysis of high-resolution resonant inelastic x-ray scattering (RIXS) spectra of 2LNO/2LTO along the $(1,0,0)_{pc=pseudocubic}$ direction. (a) Schematics of the RIXS experimental geometry. The $\theta$ ($\alpha$), $\mathbf{k}_{in}$ ($\mathbf{k}_{out}$), and $\mathbf{q}$ are the incident (scattering) angle, the incoming (outgoing) momentum, and the momentum transfer, respectively. (b) – (i) Fitting analysis of the high-resolution RIXS spectra of 2LNO/2LTO along the $(1,0,0)_{pc}$ direction (expressed in reciprocal lattice units of $|\mathbf{q}_{100}| \equiv 1.60$ Å$^{-1}$). The raw data are first self-absorption corrected (SAC) and then fitted using four components: the elastic peak, a low-energy phonon, a dominant magnon, and electron-hole (e-h) pair excitations at higher energies.

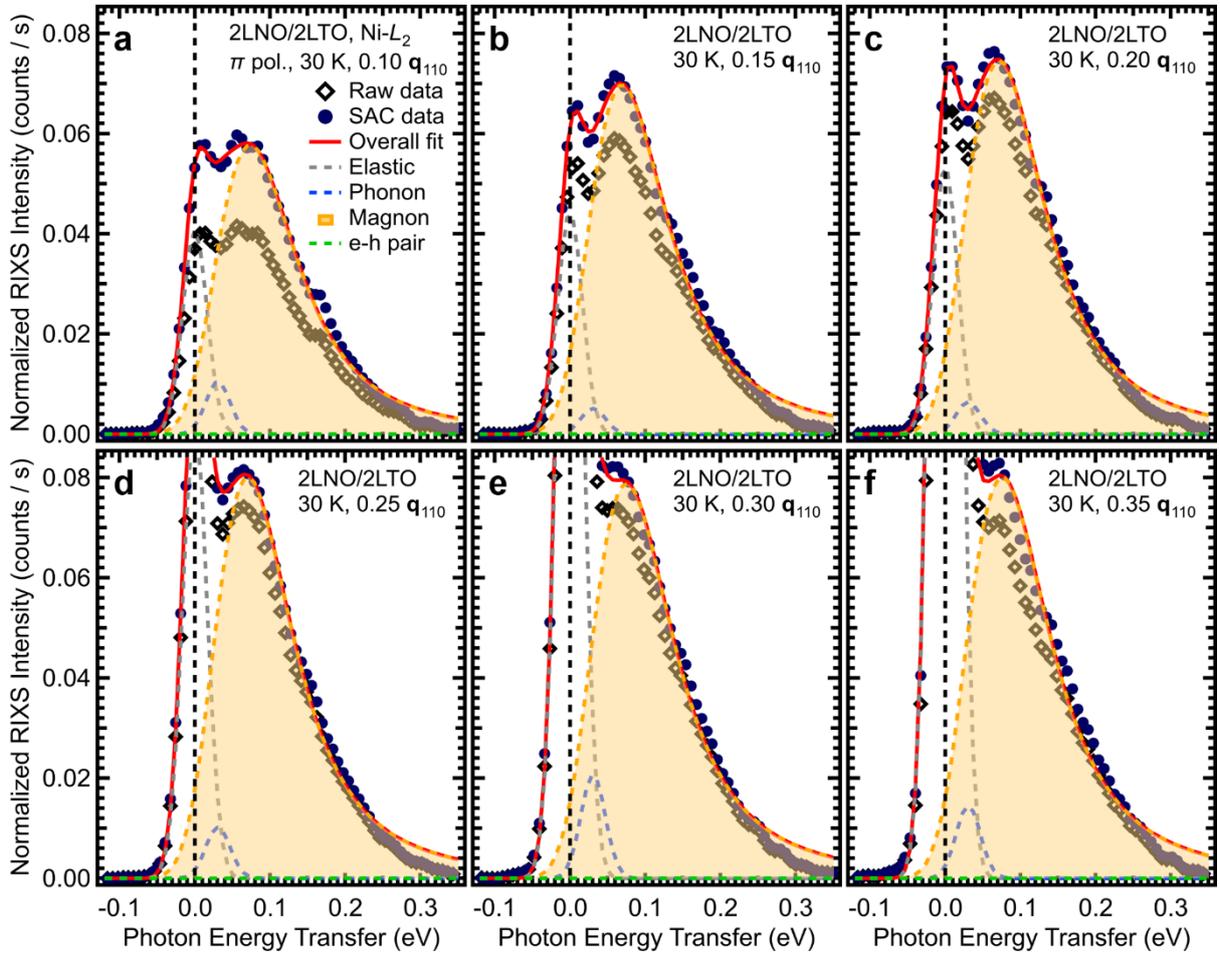

**Figure S7.** Fitting analysis of high-resolution resonant inelastic x-ray scattering (RIXS) spectra of 2LNO/2LTO along the $(1,1,0)_{pseudocubic}$ direction. The momentum transfer is expressed in reciprocal lattice units of $|\mathbf{q}_{110}| \equiv 2.26$ Å$^{-1}$. The raw data are first self-absorption corrected (SAC) and then fitted using four components: the elastic peak, a low-energy phonon, a dominant magnon, and electron-hole (e-h) pair excitations at higher energies.

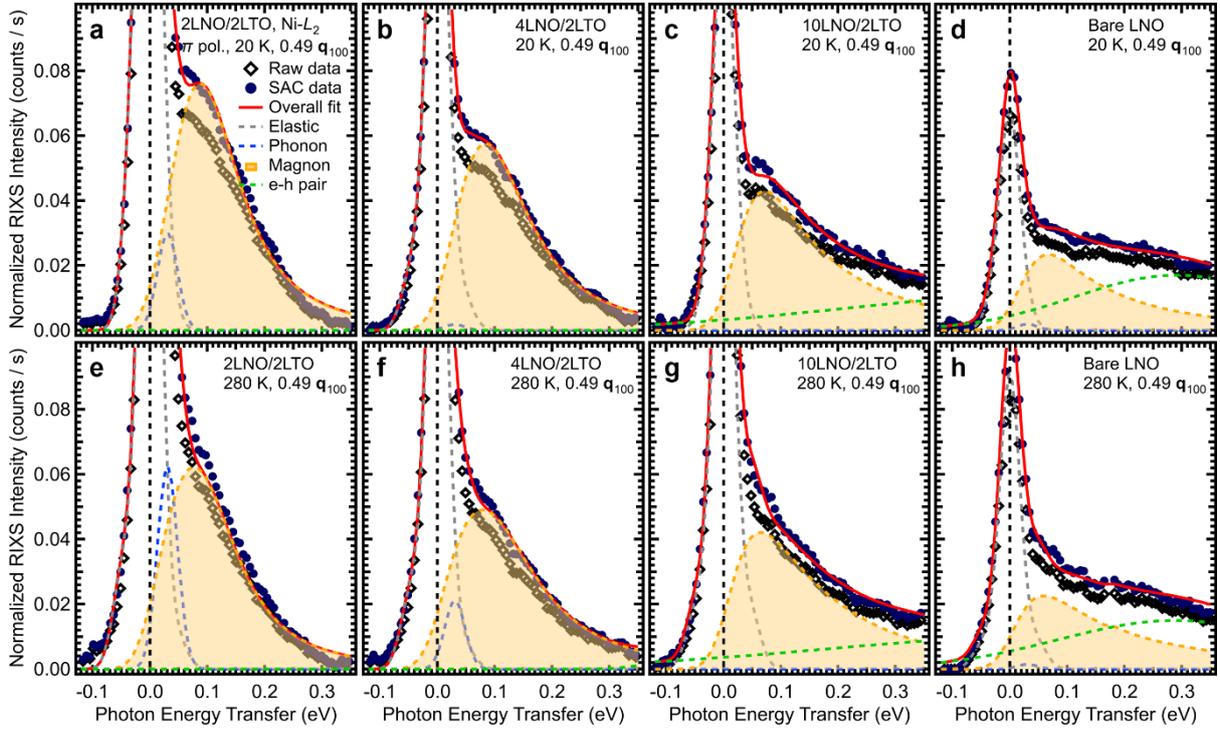

**Figure S8.** Fitting analysis of temperature-dependent high-resolution resonant inelastic x-ray scattering (RIXS) spectra of LNO/LTO superlattices. The analysis compares the high-resolution RIXS spectra of different LNO/LTO superlattices with those of the bare LNO sample at two different temperatures: 20 K and 280 K. The spectra are taken along the $(1,0,0)_{pseudocubic}$ momentum direction, expressed in reciprocal lattice units of $|\mathbf{q}_{110}| \equiv 2.26$ Å$^{-1}$. The raw data are first self-absorption corrected (SAC) and then fitted using four components: the elastic peak, a low-energy phonon, a dominant magnon, and electron-hole (e-h) pair excitations at higher energies.

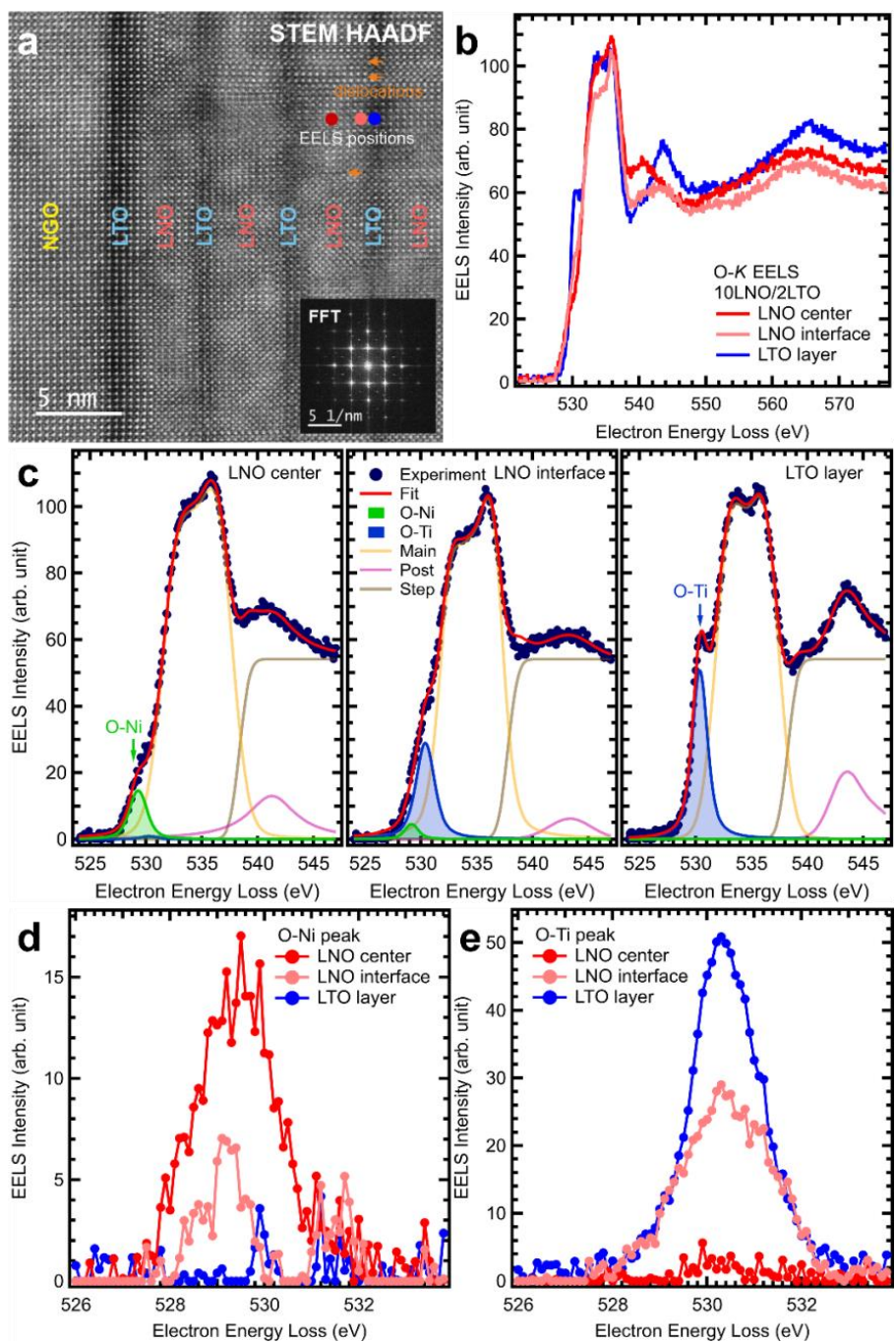

**Figure S9.** Scanning transmission electron microscopy (STEM) and cross-sectional electron energy loss spectroscopy (EELS) of 10LNO/2LTO. (a) High-resolution high-angle annular dark-field (HAADF) STEM image of 10LNO/2LTO. The inset shows the corresponding Fast-Fourier-Transform (FFT) demonstrating its crystal order and [100] orientation. (b) Cross-sectional O-*K* EELS spectra of 10LNO/2LTO taken from points representatively denoted by filled circles in (a) with the corresponding colour. (c) Fitting analysis of the EELS spectra to disentangle the spectral weights of O-Ni and O-Ti peaks from the other O-*K* peaks. (d) and (e) Disentangled O-Ni and O-Ti peaks, respectively, after the spectral weight from all the other peaks have been subtracted.